\documentclass{IEEEtran}
\usepackage{cite}
\usepackage{amsmath,amssymb,amsfonts}
\usepackage{algorithmic}
\usepackage{hyperref}
\usepackage{url}
\usepackage{graphicx}
\usepackage{booktabs}
\usepackage{array}
\usepackage{textcomp}
\usepackage{subcaption}
 \usepackage{ragged2e} 
\usepackage{xcolor}
\usepackage{multirow}

\usepackage{flushend}  
\usepackage{cuted}    
\usepackage{braket}

\begin{document}

\title{Three-Qubit Quantum Energy Teleportation Protocol for Significantly High Energy Efficiency Utilizing Superconducting Qubits}

\author{
\IEEEauthorblockN{
Md~Shoyib~Hassan\IEEEauthorrefmark{1},
Golam~Dastegir~Al~Quaderi\IEEEauthorrefmark{2},
M.~R.~C.~Mahdy\IEEEauthorrefmark{1,}\IEEEauthorrefmark{3,}\IEEEauthorrefmark{4}
}\\
\IEEEauthorblockA{
\IEEEauthorrefmark{1}Department of Electrical and Computer Engineering\\
North South University, Dhaka, Bangladesh
}\\
\IEEEauthorblockA{
\IEEEauthorrefmark{2}Department of Physics, University of Dhaka, Dhaka, Bangladesh
}\\
\IEEEauthorblockA{
\IEEEauthorrefmark{3}NSU Center of Quantum Computing, North South University, Dhaka, Bangladesh\\
\IEEEauthorrefmark{4}Corresponding Author: mahdy.chowdhury@northsouth.edu
}
}

\maketitle

\begin{abstract}
Quantum Energy Teleportation (QET) is a novel method that leverages quantum entanglement to transfer energy between two distant locations without any physical movement of the energy. The first realization of QET on superconducting hardware, utilizing a 2-qubit system, demonstrated energy retrieval at the receiver, Bob, corresponding to a work efficiency of about $11.7\%$ (extracted work relative to the energy deposited by the sender). In this work, for the very  first time, we have presented a completely new approach using a 3-qubit system to enhance the energy efficiency of QET. We have incorporated a novel 3-qubit ground state Ising Model Hamiltonian $H$ to achieve this, which conforms to the constraints of Zero mean energy and anti-commutative properties of the operations on the observable of the senders and receiver. Our experimental results show a significant improvement in terms of energy retrieval. We study two protocols. In the Single-Input Multiple-Output (SIMO) model, a single sender (Alice) deposits energy $E_0$ and two receivers (Bob and Charlie) jointly extract negative energy; with every interaction term retained, including the inter-receiver bond, the honest efficiency is $\eta = W/E_0 \approx 8\%$ to $10\%$, comparable to the $\approx 11.7\%$ of the minimal 2-qubit model but now distributed over two outputs from a single deposit. In the Multiple-Input Single-Output (MISO) model, two senders (Alice and Charlie) deposit a combined energy through a single entangle-then-measure measurement and a single receiver (Bob) extracts it; after removing the energy the entangler places directly into the receiver's bond, the net teleportation efficiency relative to the senders' deposit is $34\%$ to $42\%$, well above the 2-qubit value, with energy conservation respected. In both, energy injected locally into a quantum many-body ground state is extracted at a distant location, the negative energy density at the receiver being the signature of the protocol. Consequently, our novel Ising Model Hamiltonian is based on 3-qubit time-evolution energy dynamics that enables exploring non-trivial topological characteristics, robust fault tolerance and a better approximations of quantum fields. This achievement not only marks a step forward in practical quantum energy applications but also provides a new framework for future research in quantum energy teleportation. Given that numerous technologies have already adopted the QET protocol, researchers can now integrate this enhanced protocol into existing systems for improved functionality.\\
Code : https://github.com/C0dE-l3eAkeR/3-Qubit-QET/

\end{abstract}


\section{Quantum Energy Teleportation (QET)}

Alongside the fact that information about quantum state teleportation to distant locations is widely recognized \cite{key1,key2,key3,key4}, it is also as widely understood that quantum state energy can be transmitted similarly, paving the potential for future usage. Quantum information transmitted through quantum teleportation is intangible, whereas energy is clearly defined as a measure of physical quantity. Transmitting physical quantities to distant locations was a somewhat uncharted domain of technology before Quantum Energy Teleportation (QET) was first theoretically suggested by Hotta approximately 17 years ago. Since then, it has been the subject of theoretical investigation in spin chains \cite{key5,key6,key7}, a quantum Hall system \cite{key8}, an ion trap system\cite{key9}, and other diverse systems \cite{key10,key11,key12} that are still at a theoretical level. Surprisingly, the experimental validation of QET has been infrequent before the work described in \cite{key13}, despite its feasibility and scalability with a relatively simple quantum system. The initial empirical validation of Quantum Energy Teleportation (QET) using real cloud-based quantum computers has been conducted in \cite{key14} in a very prominent manner, with the necessary quantum circuits to do this. They successfully implemented Quantum Energy Teleportation (QET) on the IBM quantum environment that leverages superconducting quantum computers by employing quantum error mitigation techniques \cite{key15,key16,key17}. But the problem lays in the efficiency in terms of energy gain. Their experimental results indicate that we can extract only a scant amount of the total energy teleported by the sender. 

Though this paper is the first one to explore energy teleportation protocol employing a 3-qubit system, several works have been done on the 3-qubit information teleportation. In \cite{key38} author investigates the impact of noise on quantum teleportation using GHZ and non-standard W states, showing that the W state maintains higher fidelity over time. The study highlights that weak and reverse measurements do not significantly enhance teleportation efficiency in noisy conditions. Subsequently, in \cite{key39} the authors introduce eight GHZ-like states that enable both standard and controlled teleportation with perfect fidelity using a "magic bases" framework. The work allows for flexible qubit distribution among parties, offering a highly efficient and adaptable teleportation protocol. Finally, in \cite{key40} author demonstrates that a specific three-qubit state can be teleported using a simplified four-qubit entangled state, reducing the complexity of the protocol. By introducing one ancillary qubit and CNOT operations, the authors present a more practical and feasible teleportation scheme for multi-qubit systems.

\begin{figure*}[!t]
    \centering
    \includegraphics[width=0.85\textwidth]{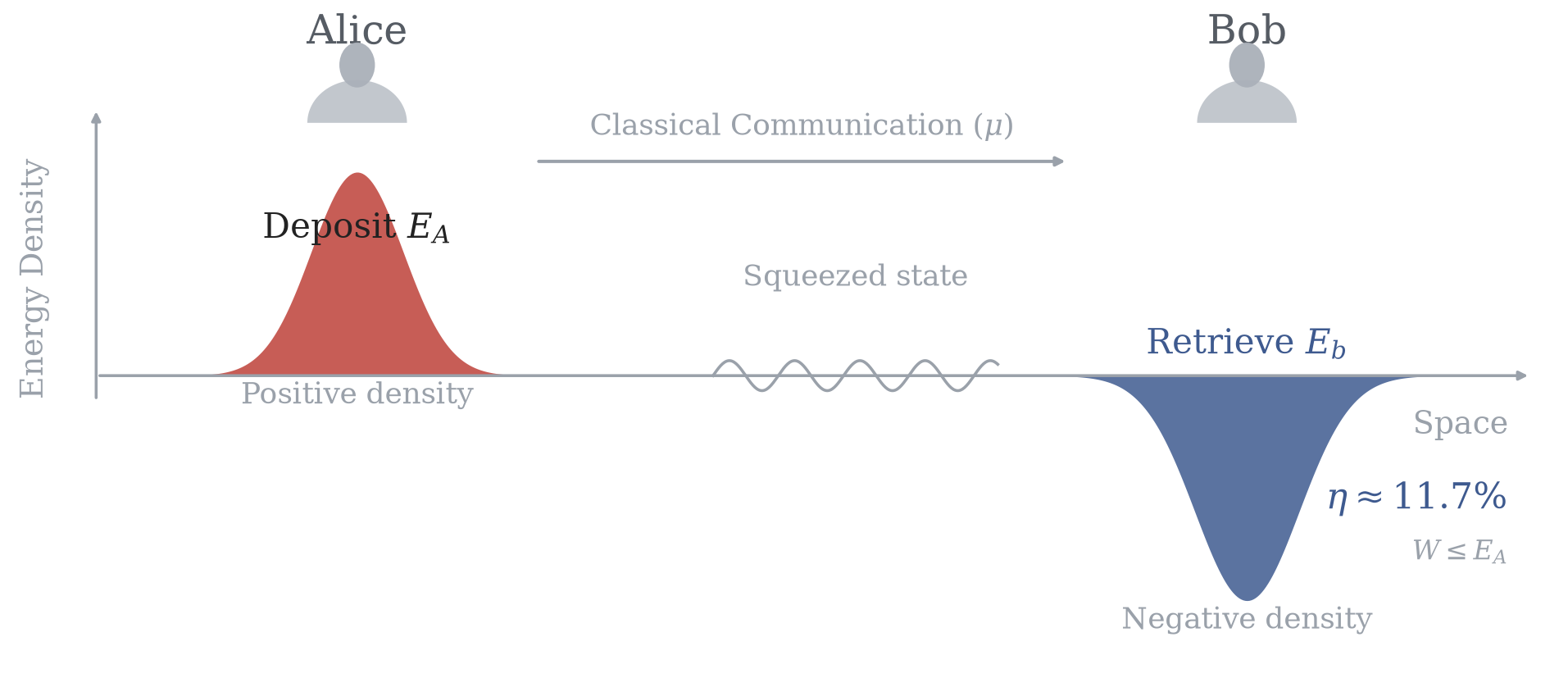}
   \caption{Energy-density schematic of the minimal 2-qubit QET. Alice deposits energy $E_A$ into the ground state (positive energy density); after classical communication of her measurement outcome $\mu$, Bob extracts negative energy density at a distant location, with work efficiency $\eta\approx 11.7\%$ and energy conservation respected ($W\le E_A$).}
   \label{fig:minimal_schema}
\end{figure*}

\begin{figure*}[!t]
    \centering
    \begin{subfigure}[b]{0.49\textwidth}
        \includegraphics[width=\textwidth]{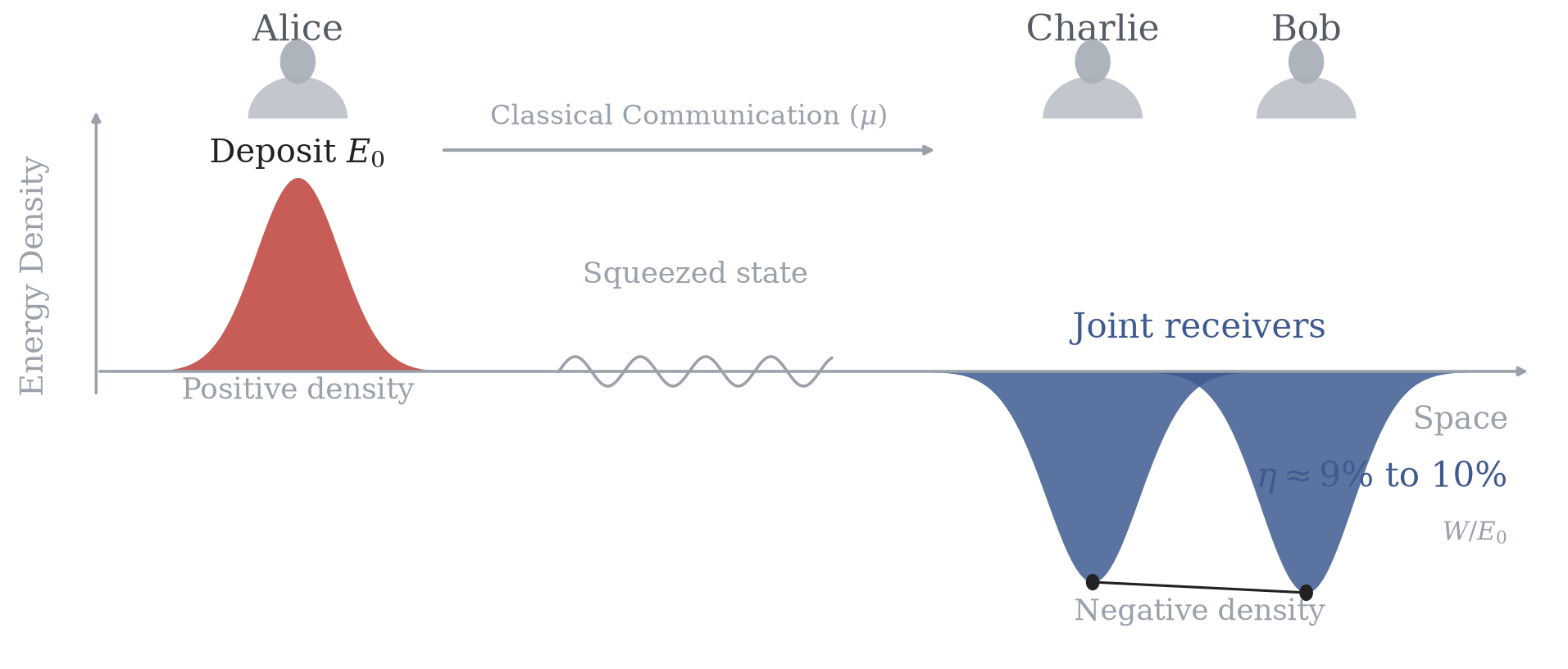}
        \caption{SIMO: Alice deposits $E_0$; the two receivers Bob and Charlie jointly extract negative energy, with efficiency $\eta\approx 8\%$ to $10\%$ (the inter-receiver bond $V_{1,2}$ fully included).}
    \end{subfigure}
    \hfill
    \begin{subfigure}[b]{0.49\textwidth}
        \includegraphics[width=\textwidth]{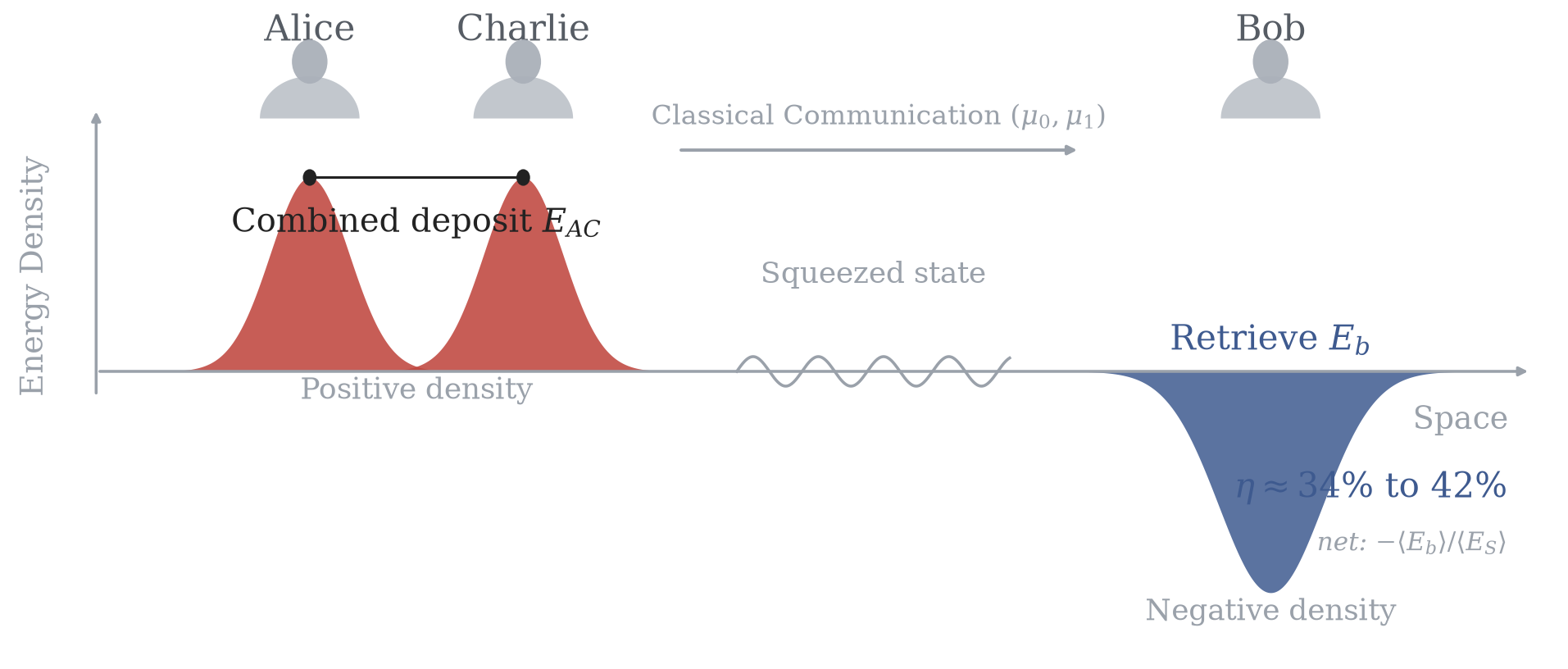}
        \caption{MISO: Alice and Charlie deposit a combined energy $E_{AC}$ through a joint (entangle-then-measure) measurement; Bob retrieves negative energy with net efficiency $\eta\approx 34\%$ to $42\%$, energy conservation respected.}
    \end{subfigure}
    \caption{Energy-density schematics of the two protocols.}
    \label{fig:QET}
\end{figure*}
The purpose of this paper is to ameliorate the efficiency of the energy retrieved from this protocol by extending the number of qubits used. The three quantum hardware utilized in our study is IBM's quantum computer, ibm\_brisbane, ibm\_kyv and ibm\_sherbrooke, which are easily accessible to everyone worldwide at zero cost. By utilizing the quantum circuits presented in this research, individuals will have the capability to replicate the outcomes and quantum energy teleportation (QETs) efficiently. Given that the features of quantum computers are openly accessible nowadays, it will be feasible for everyone to use the extended QET protocol. The techniques we have introduced enhancing the minimal QET model \cite{key14} can be utilized on any system that use QET for energy transfer.

The following explanation clarifies why QET serves as a universal method for quantum energy transfer, analogous to how quantum teleportation functions as a universal method for transferring quantum information. Excited states are indigenous to the observations of the ground state of a quantum many-body system, which subsequently raises the expected energy level. It should be noted that the experimental devices provide the additional energy. The ground state of a quantum many-body system possesses the significant characteristic of entanglement, which results in quantum fluctuations in the overall energy of the ground state. To clarify, the energy fluctuations of the local systems are entangled due to quantum effects. Measuring subsystem A at the local level, results in the destruction of the entanglement of the ground state. Similarly, this statement is true for any other subsystems C entangled to the whole system. The measurement instrument transfers energy $E_A$, $E_C$ into the entire system. The injected energy $E_A$ and $E_C$  remains localized within subsystems A and C throughout the initial phases of time evolution. However, activities focused solely on subsystems A or C cannot withdraw $E_A$ or $E_C$ from the system. This is because information about $E_A$ and $E_C$ is distributed across remote locations in addition to A and C, due to the pre-existing entanglement. In other words, the energy injected locally, denoted as $E_A$ and $E_C$, can be partially retrieved from any point other than A and C \cite{key18}. The QET protocol enables this capability through measurements of the ground state energy of local and semi-local Hamiltonian. As this is the key characteristic of QET, that is its complete realization through the inherent properties of the quantum many-body system ground state and the universally observed phenomenon called measurement and We observed that incorporating one extra entangled qubit increases the inherent interaction energies, which in turn allows us to enhance the protocol's efficiency.

The QET model described in \cite{key14} is a simple implementation that employs real quantum networks and quantum computers in a quantum circuit. However, the limitation resides in the efficiency of the teleportation process: on a work basis the 2-qubit model recovers only about $11.7\%$ of the energy that Alice initially deposits (extracted work relative to deposited energy). This research utilized an expanded iteration of the aforementioned method, employing quantum circuits consisting of three qubits for above mentioned Quantum Energy Teleportation (QET), as illustrated in Fig.~\ref{fig:QET}. Quantum computers already possess sufficient capability to execute a circuit depth more than 6.

Our enhanced 3-qubit Quantum Energy Teleportation (QET) framework utilizes two distinct models: SIMO (Single Input, Multiple Output) and MISO (Multiple Input, Single Output). In the SIMO model, a single sender (Alice) deposits energy and two receivers (Bob and Charlie) jointly extract it; with every interaction term retained, including the inter-receiver bond, the honest efficiency is $\eta = W/E_0 \approx 8\%$ to $10\%$. In the MISO model, two senders deposit a combined energy through a single entangle-then-measure measurement and a single receiver extracts it, with a net teleportation efficiency of $34\%$ to $42\%$ (the negative energy the receiver extracts, after removing the energy the entangler deposits directly into his bond, relative to the senders' deposit; energy conservation respected). In both cases, energy injected locally into a quantum many-body ground state is partially extracted at a distant location. Critically, this "negative energy" extraction depends on pre-existing entanglement within local and semi-local Hamiltonians. Our novel Ising Model Hamiltonian, constructed around 3-qubit time-evolution energy dynamics, governs the energy spectrum, unfolding previously unexplored energy eigenstates. These states exhibit unique properties, such as exotic forms of entanglement that allows us to employ non-trivial topological characteristics (e.g. MISO, SIMO), potentially advancing our understanding of quantum matter. Additionally, it enables the design of more robust quantum gate simulations with higher fault tolerance that potentially reduced the error due to qubit decoherence and noise. While the experimental realization uses 3 qubits, the theoretical formulation of the Hamiltonian is derived from lattice field theory principles. It serves as a minimal lattice discretization that preserves the essential symmetries, renormalization group properties, and vacuum entanglement structures of the continuous Quantum Field Theory. Thus, our Ising Model Hamiltonian bridges the gap between quantum mechanics and quantum field theory by providing a lattice model that better mimics specific quantum fields approximation. 

Our work presents theoretical advancements with significant implications for condensed matter and quantum field theories. Our Hamiltonian's design, based on 3-qubit energy dynamics has enabled us to incorporate an exotic form of entanglement that has been induced from Zero
energy eignestates. This eigenstate facilitates the use of a lattice model like SIMO, to extract
negative ground state energy through a process we term Metrotropy. Metrotropy refers to the
maximal amount of energy that can be extracted from a finite quantum system using projective
measurements, rather than the unitary operations used in traditional ergotropy. The incorporation
of the SIMO lattice model provides a closer approximation to specific quantum fields and
efficient renormalization techniques. Renormalization, as applied in quantum field theory, is the
process of refining models to account for effects at different scales, such as filtering out high energy fluctuations. It strengthens fidelity maintenance in simulations, especially in many-body
quantum systems where entanglement and interaction energies $\langle V \rangle$ play critical roles. When
projecting a quantum state onto a new basis (as part of energy extraction), the transition
probabilities form a bistochastic matrix, which governs the possible final distributions of energy
levels after measurement. A bistochastic matrix (also called a doubly stochastic matrix) is a
square matrix where all elements are non-negative, and each row and column sums to 1. In our
work, the exact permutation of using rotations (e.g., Ry or Rz gates) to the qubits before
measurement effectively select an optimal basis for the measurement. In this way, our depicted
lattice model SIMO allows us to derive a bistochastic matrix that ensures that the sender’s
measurement collapses the entangled ground state, redistributing a part of the system's energy
that becomes accessible to a distant receiver. A local measurement performed at a second site can then be fed forward to enhance this extraction at no additional energy cost.
Furthermore, topologically ordered states in our Hamiltonian rely on enhanced symmetry, where
the global properties of the state are resistant to local errors. Enhanced symmetry groups in a
Hamiltonian, mean that the system’s energy levels, or ground states are more stable and less
sensitive to certain types of perturbations or noise. Consequently, designing quantum gates
within systems that have enhanced symmetry groups, leverage these symmetries to maintain
coherence and reduce errors enabling us to gain maximum efficiency in terms of energy
extraction.

\subsection{Defining The Essentials Of QET}

\begin{figure*}[!t]
\centering
    \begin{minipage}[t]{0.49\textwidth}\centering\includegraphics[width=\textwidth]{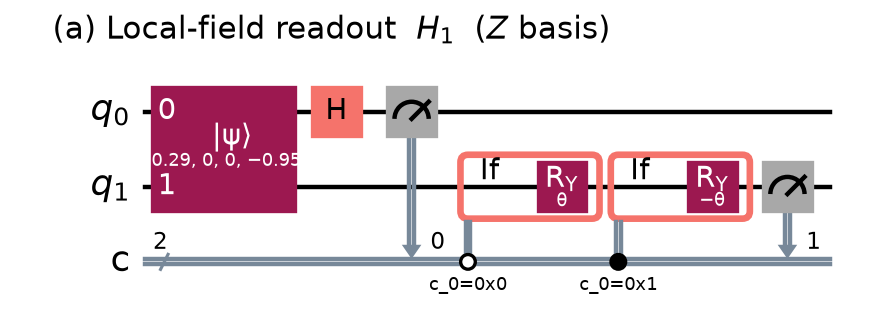}\end{minipage}\hfill
    \begin{minipage}[t]{0.49\textwidth}\centering\includegraphics[width=\textwidth]{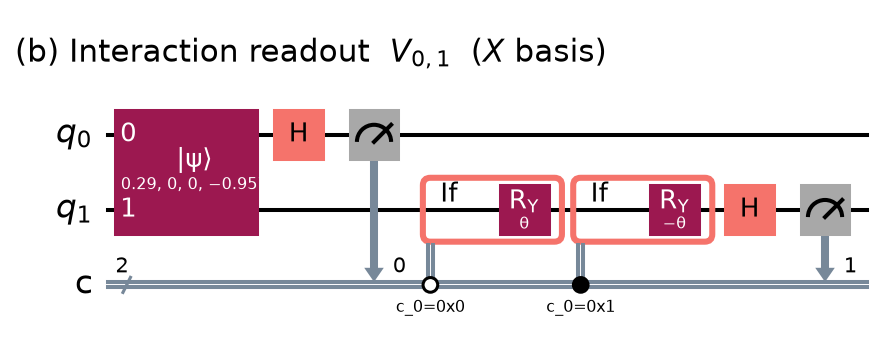}\end{minipage}
    \caption{Quantum circuit for the minimal 2-qubit QET (Alice $=q_0$, Bob $=q_1$). Alice measures $X_0$ (a Hadamard followed by a computational-basis measurement); conditioned on her outcome $\mu$, Bob applies the rotation $R_Y(\pm\theta)$ and reads his local field $H_1$ (panel a, left, $Z$ basis) and his interaction with Alice $V_{0,1}$ (panel b, right, $X$ basis).}
    \label{fig:example}
\end{figure*}

 To begin, we will provide a comprehensive overview of the QET protocol\cite{key18}. To find quantum circuit implementations for specific situations, refer to \cite{key13,key14,key20,key21} and Fig.~\ref{fig:example}, where the local Hamiltonian $H = \sum_{n=0}^N H_n$ is defined and $H_n$ represents the local Hamiltonian that interacts with surrounding qubits. It must satisfy the following  constraints
\begin{equation}
\langle g| H |g\rangle = \langle g| H_n |g\rangle = 0, \quad \forall n \in \{1, \cdots , N\},
\end{equation}

Here $|g\rangle$ is said to be the ground state of the total Hamiltonian $H$. But in case of local $H_n$ it might always not be the case. It is crucial to acknowledge that  $|g\rangle$  is a state of entanglement in a generic context. To uphold the requirement (1), it is possible to consistently sum or deduct constant values, As the ground state, it is evident that any non-trivial (local) operations to $|g\rangle$ such as measurement, results in increase of the energy expectation value.

Below, we provide a description of the QET protocol. Alice plays the role of energy supplier while Bob remains as a receiver. Alice does a projective measurement on her Pauli operator $\sigma_A$, using operator $P_A(\mu) = \frac{1}{2} ( 1 + \mu \sigma_A )$. The results she obtains is either $\mu = -1$ or $\mu = +1$.  $E_A$ that is the injected energy is localized around subsystem $A$, but Alice is unable to withdraw it from the system merely through her operations at $A$. Nevertheless, by employing LOCC, Bob has the ability to extract a certain amount of energy from his local system.

Alice transmits her measurement result  $\mu$ to Bob by classical communication. Upon receiving the result, Bob applies conditional operation $U_b(\mu)$ to his state and performs a measurement on his local Hamiltonian $H_B$. His operation can be defined as
\begin{equation}
U_b(\mu) = \cos \theta I - i \mu \sin \theta \sigma_B,
\end{equation}
where $\theta$ obeys
\begin{equation}
\cos(2\theta) = \frac{\xi}{\sqrt{\xi^2 + \eta^2}}, \sin(2\theta) = - \frac{\eta}{\sqrt{\xi^2 + \eta^2}}
\end{equation}
where
\begin{equation}
\xi = \langle g| \sigma_B H \sigma_B |g\rangle, \quad \eta = \langle g| \sigma_A \dot{\sigma}_B |g\rangle,
\end{equation}
Here $\dot{\sigma}_B = i [H_b, \sigma_B] = [H, \sigma_B]$ must be maintained by the local hamiltonian. The average quantum state $\rho_{\text{QET}}$, additionally a mixed state, can be gained after applying Bob's operator $U_b(\mu)$ to $\frac{1}{\sqrt{p(\mu)}} P_A (\mu) |g\rangle$, where $p(\mu)$ is depicted as a normalization factor.

We get the density matrix $\rho_{\text{QET}}$ after Bob applies the operator $U_b(\mu)$ to $P_A (\mu) |g\rangle$ is
\begin{equation}
\rho_{\text{QET}} = \sum_{\mu \in \{ \pm 1 \}} U_b(\mu) P_A (\mu) |g\rangle \langle g| P_A (\mu) U_b^{\dagger} (\mu).
\end{equation}
 Bob's expected energy at his local system can be measured as
\begin{equation}
\langle E_b \rangle = \text{Tr}[\rho_{\text{QET}} H_B] = \frac{1}{2} \left[ \xi - \sqrt{\xi^2 + \eta^2} \right],
\end{equation}
which is evaluated negative if $\eta \neq 0$ and no energy dissipation, the positive energy of $- \langle E_b \rangle$ is teleported to Bobs device by the law of energy conservation.

\subsection{Minimal QET Model}

For the full description of the minimal model, schematically depicted in Fig.~\ref{fig:minimal_schema}, refer to \cite{key14}.
Let's assume $k, h$ be positive numbers. The minimal model is defined as
\begin{align}
    H_{\text{tot}} &= H_0 + H_1 + V, \\
    H_n &= h Z_n + \frac{h^2}{\sqrt{h^2 + k^2}}, \quad (n = 0, 1) \\
    V &= 2k X_0 X_1 + \frac{2k^2}{\sqrt{h^2 + k^2}}. \label{eq:minimal_V}
\end{align}
The ground state of $H_{\text{tot}}$ can be defined as
\begin{equation}
    | g \rangle = \frac{1}{\sqrt{2}} \sqrt{1 - \frac{h}{\sqrt{h^2 + k^2}}} |00\rangle - \frac{1}{\sqrt{2}} \sqrt{1 + \frac{h}{\sqrt{h^2 + k^2}}} |11\rangle,
\end{equation}
One can add the constant terms to the Hamiltonians so that the ground state $| g \rangle$ of $H_{\text{tot}}$ evaluates as zero mean energy for all local and global Hamiltonians:
\begin{equation}
    \langle g | H_{\text{tot}} | g \rangle = \langle g | H_0 | g \rangle = \langle g | H_1 | g \rangle = \langle g | V | g \rangle = 0.
\end{equation}
As we discussed earlier, $| g \rangle$ can be neither a ground state nor an eigenstate of $H_n, V, H_n + V$, where $\{n = 0, 1\}$. The primary objective seems to be obtaining negative ground state energy of local and semi-local Hamiltonians from QET protocol.

The QET protocol is outlined below. Alice initially performs a measurement on her Pauli operator $X_0$ by $P_0 (\mu) = \frac{1}{2} (1 + \mu X_0)$ resulting in $\mu = -1$ or $+1$. t Alice’s expectation energy is denoted as,
\begin{equation}
    E_0 = -\frac{h^2}{\sqrt{h^2 + k^2}}.
\end{equation}
She communicates her measurement result $\mu$ to Bob using a classical channel, who conducts an operation $U_1 (\mu)$ to his qubit and measures $H_1$ and $V$. Bobs given by the following:
\begin{equation}
U_1(\mu) = \cos\phi \, I - i \mu \sin\phi \, Y_1 = R_Y(2\phi)
\end{equation}
where $0 \leq \phi \leq \pi /2$ obeys
\begin{align}
    \sin(2\phi) &= \frac{hk}{\sqrt{(h^2 + 2k^2)^2 + h^2 k^2}}. 
\end{align}

The density matrix $\rho_{\text{QET}}$ is evaluated after Bob operates $U_1 (\mu)$ to $P_0 (\mu) | g \rangle$ as
\begin{equation}
    \rho_{\text{QET}} = \sum_{\mu \epsilon \{-1, 1\}} U_1 (\mu) P_0 (\mu) | g \rangle \langle g | P_0 (\mu) U_1^{\dagger} (\mu).
\end{equation}
By using $\rho_{\text{QET}}$, the expected local energy at Bob’s subsystem is calculated as $\langle E_1 \rangle = \text{Tr} [\rho_{\text{QET}} (H_1 + V)]$, which comes out negative in general. By the law of energy conservation, $E_b = - \langle E_1 \rangle ( > 0)$ is extracted from the system by the device that operates $U_1 (\mu)$ \cite{key23}.\\

\section{Quantum Circuit Implementation of Extended QET Model}

We leverage the capabilities of the 3-qubit QET model through two distinctive configurations: the Multiple-Input Single-Output (MISO) model and the Single-Input Multiple-Output (SIMO) model. In this section, we first construct the 3-qubit Hamiltonian and its entangled ground state, which are common to both configurations, and then discuss each protocol in detail.

\subsection{The 3-Qubit Hamiltonian and Ground State}

\subsubsection{Defining the 3-Qubit Hamiltonian}

We define a novel 3-qubit Hamiltonian $H_{\text{tot}}$, where $h$ and $k$ are positive real parameters analogous to those in the minimal qubit model. The local and interaction Hamiltonians are constructed as follows:

\begin{align}
    H_n &= h Z_n + c_H , \quad (n = 0, 1, 2) \label{eq:Hn_def} \\
    V_{i,j} &= k X_i X_j + c_V , \label{eq:Vij_def}\\
    H_{\text{tot}} &= H_0 + H_1 + H_2 + V_{0,1} + V_{1,2} + V_{0,2}, \label{eq:Htot_def}
\end{align}

The scalar constants $c_H$ and $c_V$ in Eqs.~\eqref{eq:Hn_def} and \eqref{eq:Vij_def} are fixed so that the ground state $|g\rangle$ of $H_{\text{tot}}$ yields zero mean energy for all local, semi-local, and global Hamiltonians (their closed-form values are given in Eq.~\eqref{eq:constants} below). This constraint is fundamental to the QET protocol and is expressed mathematically as:

\begin{equation}
\begin{aligned}
\langle g|H_{\text{tot}}|g\rangle &= \langle g|H_0|g\rangle = \langle g|H_1|g\rangle = \langle g|H_2|g\rangle \\
&= \langle g|V_{0,1}|g\rangle = \langle g|V_{1,2}|g\rangle = \langle g|V_{0,2}|g\rangle = 0.
\end{aligned}
\label{eq:zero_energy_all}
\end{equation}

Consequently, for the 3-qubit system, the total Hamiltonian simplifies to:

\[
\begin{aligned}
H_{\text{tot}} = {}& h\,(Z_0 + Z_1 + Z_2)
+ k\,(X_0 X_1 + X_1 X_2 + X_0 X_2) \\
& + \bigl(2\sqrt{h^{2} + hk + k^{2}} + h - k\bigr)\, \mathbb{I}.
\end{aligned}
\]

The ground state of $H_{\text{tot}}$, which is the lowest energy state of this Hamiltonian. We can get this by taking the corresponding eigen vector of the lowest eigen value of $H_{\text{tot}}$, which lies in the odd-parity subspace spanned by $\{|001\rangle, |010\rangle, |100\rangle, |111\rangle\}$ and, in properly normalized form, reads:
\begin{equation}
    |g\rangle = \frac{1}{\mathcal{N}}\bigl(|111\rangle - M\,(|100\rangle + |010\rangle + |001\rangle)\bigr),
    \label{eq:ground_state}
\end{equation}
with normalization $\mathcal{N} = \sqrt{1 + 3M^2}$.

The auxiliary parameter is
\begin{equation}
K = \sqrt{h^2 + hk + k^2}. \label{eq:auxK}
\end{equation}
Although a direct symbolic diagonalization of $H_{\text{tot}}$ returns three apparently distinct rational functions of $h$ and $k$ for the coefficients of $|100\rangle$, $|010\rangle$, and $|001\rangle$, the full $S_3$ permutation symmetry of the Hamiltonian among the three qubits maps these basis states into one another and thereby forces their coefficients to coincide. The non-degenerate ground state lies in the symmetric sector, and the three coefficients collapse to a single closed form,
\begin{equation}
M_1 = M_2 = M_3 \equiv M = \frac{k}{2h + k + 2K},
\label{eq:M}
\end{equation}
with $K$ defined in Eq.~\eqref{eq:auxK}. We have verified by direct substitution that this compact expression reproduces, to machine precision, the unwieldy rational functions returned by symbolic diagonalization at all sampled $(h,k)$; it replaces the three cumbersome coefficients of the earlier formulation by a single parameter and renders the structure of $|g\rangle$ transparent.

Equivalently, $|g\rangle$ may be expressed as $|g\rangle = \mathcal{N}^{-1} C |\psi\rangle$, where the coefficient matrix
\begin{equation*}
C = \text{diag}\begin{pmatrix}
0 & -M & -M & 0 & -M & 0 & 0 & 1
\end{pmatrix}
\end{equation*}
acts on the uniform superposition
\begin{equation*}
\begin{aligned}
|\psi\rangle = {}& |000\rangle + |001\rangle + |010\rangle + |011\rangle \\
&+ |100\rangle + |101\rangle + |110\rangle + |111\rangle.
\end{aligned}
\end{equation*}
The unnormalized form $C|\psi\rangle$ is retained solely for notational compactness in the eigenvalue derivation; the normalized form \eqref{eq:ground_state} is used consistently in all expectation-value calculations throughout this manuscript. In particular, it coincides with Eq.~\eqref{eq:gs_normalized} of Section~\ref{sec:entropy_bound}, where the same state is analyzed from a thermodynamic perspective.

It is crucial to emphasize that the state $|g\rangle$ is \textit{neither} an eigenstate \textit{nor} a ground state of the individual components $H_n$, $V_{i,j}$, or their combinations $H_n + V_{i,j}$, where $\{n = 0, 1, 2\}$ and $\{i,j = 0, 1, 2$ with $i < j\}$. This non-trivial property is the cornerstone of the QET protocol: it enables the extraction of negative ground state energy from local and semi-local Hamiltonians, a phenomenon that arises solely due to quantum entanglement in the global ground state.

The two scalar constants in Eqs.~\eqref{eq:Hn_def} and \eqref{eq:Vij_def} are fixed uniquely and in closed form by the zero-mean-energy requirement~\eqref{eq:zero_energy_all}. Imposing $\langle g|H_n|g\rangle = 0$ and $\langle g|V_{i,j}|g\rangle = 0$ on the ground state~\eqref{eq:ground_state} yields
\begin{equation}
    c_H = \frac{h\,(1 - M^{2})}{1 + 3M^{2}}, \qquad
    c_V = \frac{2kM\,(1 - M)}{1 + 3M^{2}},
    \label{eq:constants}
\end{equation}
which together satisfy $3c_H + 3c_V = 2K + h - k$, in agreement with the total constant of $H_{\text{tot}}$. Because $|g\rangle$ and $H_{\text{tot}}$ are fully symmetric under every permutation of the three qubits, a single value of $c_H$ enforces all three local conditions $\langle g|H_n|g\rangle = 0$ and a single value of $c_V$ enforces all three semi-local conditions $\langle g|V_{i,j}|g\rangle = 0$ simultaneously. The full zero-energy requirement~\eqref{eq:zero_energy_all} is therefore satisfied identically, with no free auxiliary parameter remaining.

\subsubsection{Physical Justification of the Hamiltonian Construction}

The 3-qubit Hamiltonian is a natural and principled extension of Hotta's minimal 2-qubit QET model, guided by three physical requirements essential to any valid QET protocol.

\textbf{(i) Non-commutativity of local and interaction Hamiltonians.}
The fundamental mechanism of QET relies on the fact that the local Hamiltonian $H_n$ and the interaction Hamiltonian $V_{i,j}$ do not commute. If $[H_n, V_{i,j}] = 0$, the ground state of $H_{\text{tot}}$ would be a product state with no entanglement, rendering the QET protocol trivial. The transverse-field Ising structure
\begin{equation}
    H_n = hZ_n + c_n, \qquad V_{i,j} = kX_iX_j + c_{i,j},
\end{equation}
is the minimal operator choice that guarantees $[Z_n, X_iX_j] \neq 0$ for all relevant index pairs, thereby forcing the ground state into a non-trivially entangled configuration. Intuitively, the $Z_n$ term acts as a local energy trap that favors computational-basis configurations, whereas the $X_i X_j$ interaction induces quantum fluctuations that compete with this preference. This interplay gives rise to the ground-state entanglement that serves as the resource for QET.

\textbf{(ii) Full permutation symmetry among all qubit pairs.}
In the 3-qubit setting there are three distinct qubit pairs: $(0,1)$, $(1,2)$, and $(0,2)$. We include pairwise interaction terms for all three pairs,
\begin{equation}
    H_{\text{tot}} = \sum_{n=0}^{2} H_n + V_{0,1} + V_{1,2} + V_{0,2},
\end{equation}
rather than restricting to nearest-neighbor coupling alone. We made this choice being motivated by the requirement that the zero-mean-energy constraint $\langle g|V_{i,j}|g\rangle = 0$ must be satisfied simultaneously for every qubit pair $(i,j)$. A nearest-neighbor-only coupling would break the symmetry between sender-receiver pairs, making it impossible to enforce this condition globally. Furthermore, the energy scale of the 3-qubit system is governed by the characteristic eigenvalue
\begin{equation}
    K = \sqrt{h^2 + hk + k^2},
\end{equation}
which replaces $\sqrt{h^2+k^2}$ of the 2-qubit case. The cross term $hk$ inside the square root arises naturally from the simultaneous presence of three pairwise couplings, each of which involves two sites already subject to a local $Z$ field, and appears directly in the characteristic polynomial of the Hamiltonian when restricted to the ground-state subspace $\{|001\rangle,|010\rangle,|100\rangle,|111\rangle\}$.

\textbf{(iii) Local passivity of the ground state.}
A fundamental requirement of the QET protocol is that no energy can be extracted from the ground state by any local unitary operation alone, i.e., the ground state must be locally passive in the sense of Pusz and Woronowicz \cite{PuszWoronowicz}, as formalized for finite quantum systems in \cite{FreyFunoHotta}. This is enforced by the zero-mean-energy condition \eqref{eq:zero_energy_all}. The constant shift terms in $H_n$ and $V_{i,j}$ are deliberately constructed to satisfy this condition. Without these shifts, the ground state would carry non-zero local energy expectation values, meaning a receiver could extract energy without any classical communication from the sender,violating the causal structure of the QET protocol.

\subsubsection{Justification for the Operator Form}

The choice of $Z_n$ as the local operator and $X_iX_j$ as the interaction operator is the unique minimal selection satisfying all three requirements above within the Pauli operator algebra. Specifically: $Z_n$ provides a local energy eigenbasis (the computational basis), ensuring that the ground state is well-defined and that local measurements in the $Z$-basis yield energy information; $X_iX_j$ is the simplest two-body operator that (a) anti-commutes with $Z_n$ on the relevant qubit, (b) preserves the $\mathbb{Z}_2$ parity symmetry $\prod_n X_n$ of the Hamiltonian, and (c) generates the odd-parity ground-state subspace $\{|001\rangle,|010\rangle,|100\rangle,|111\rangle\}$ through its action on the local vacuum. Furthermore, the commutation relations $[X_0, V_{0,2}] = 0$ and $[X_1, V_{1,2}] = 0$ guarantee that the local measurements performed by the senders do not inject energy directly into the receiver's subsystem. This property is a direct consequence of the $XX$ interaction. In contrast, interactions of the form $Z_iZ_j$ or $Y_iY_j$ generally violate these commutation relations, leading to unwanted measurement-induced disturbances at the receiver and undermining the causal consistency of the QET protocol. We also note that the deliberate exclusion of a three-body term $X_0X_1X_2$ is consistent with the minimality principle: while such a term is Hermitian and preserves the necessary commutation relations (since $X_i^2 = I$), its inclusion would add a further independent constraint to the zero-energy system~\eqref{eq:zero_energy_all}, generically overdetermining it and destroying the analytical tractability of the model. Its incorporation is, however, a natural direction for future investigation, as three-body interactions can generate genuinely tripartite (GHZ-type) entanglement potentially exceeding the efficiency of the present protocol.

\subsubsection{Generalization to Higher-Qubit Systems}
\label{sec:generalization}

The construction presented in this work admits a systematic, constructive generalization to arbitrary $N$-qubit systems ($N \ge 2$). We describe this procedure below in full generality, and subsequently discuss the computational limitations that arise as $N$ grows.

\paragraph{Constructive procedure.}
Consider the general $N$-qubit Ising Hamiltonian parameterized by positive real couplings $h, k > 0$ and an as-yet-undetermined global constant $\Lambda$:
\begin{equation}
    H_{\text{tot}}^{(N)}(h,k;\Lambda) = h\sum_{n=0}^{N-1}Z_n + k\sum_{0\le i<j\le N-1}X_iX_j + \Lambda\cdot\mathbb{I}.
    \label{eq:N_qubit_H}
\end{equation}
The operator $\Lambda\cdot\mathbb{I}$ commutes with every term in the Hamiltonian and therefore shifts the entire energy spectrum rigidly without altering the eigenstates. In particular, if $\{E_m(h,k;0)\}_{m=0}^{2^N-1}$ denotes the set of eigenvalues of $H_{\text{tot}}^{(N)}$ evaluated at $\Lambda = 0$, ordered so that $E_0 \le E_1 \le \cdots \le E_{2^N-1}$, then the eigenvalues at arbitrary $\Lambda$ satisfy
\begin{equation}
    E_m(h,k;\Lambda) = E_m(h,k;0) + \Lambda, \qquad \forall\; m \in \{0,\ldots,2^N-1\}.
    \label{eq:eigenvalue_shift}
\end{equation}
The eigenstates $\{|\psi_m\rangle\}$ remain independent of $\Lambda$.

We now exploit this uniform spectral shift to enforce the zero ground-state energy condition. Setting
\begin{equation}
    \Lambda_N \;\equiv\; -\,E_0(h,k;0),
    \label{eq:Lambda_fix}
\end{equation}
immediately yields
\begin{equation}
    E_0(h,k;\Lambda_N) = E_0(h,k;0) + \Lambda_N = 0.
    \label{eq:zero_gs}
\end{equation}
Since $E_0$ is the \emph{lowest} eigenvalue, the ground state $|g\rangle \equiv |\psi_0\rangle$ now possesses zero energy expectation, $\langle g| H_{\text{tot}}^{(N)} |g\rangle = 0$, and all excited states carry strictly positive energy $E_m > 0$ for $m \ge 1$. Consequently, the ground state is automatically the minimum-energy state of the system, fulfilling the foundational requirement of the QET protocol.

\paragraph{Subsystem decomposition and zero-mean-energy enforcement.}
With $\Lambda_N$ determined by Eq.~\eqref{eq:Lambda_fix}, the total Hamiltonian is decomposed into local and interaction terms of the form
\begin{equation}
\begin{aligned}
    H_n &= h Z_n + c_H^{(N)}, && (n = 0,\ldots,N-1), \\
    V_{i,j} &= k X_i X_j + c_V^{(N)}, && (0 \le i < j \le N-1),
\end{aligned}
    \label{eq:N_decomposition}
\end{equation}
where the subsystem constants $c_H^{(N)}$ and $c_V^{(N)}$ are fixed by demanding zero mean energy for every individual term,
\begin{equation}
    \langle g | H_n | g \rangle = 0, \qquad \langle g | V_{i,j} | g \rangle = 0, \qquad \forall\; n, \; \forall\; i < j.
    \label{eq:zero_mean_all_N}
\end{equation}
Explicitly, these conditions yield
\begin{equation}
    c_H^{(N)} = -h\,\langle g | Z_n | g \rangle, \qquad
    c_V^{(N)} = -k\,\langle g | X_i X_j | g \rangle,
    \label{eq:cH_cV_general}
\end{equation}
which are uniquely determined once the ground state $|g\rangle$ is known. If the Hamiltonian possesses the full $S_N$ permutation symmetry (uniform $h$ and $k$), a single value of $c_H^{(N)}$ enforces all $N$ local conditions and a single value of $c_V^{(N)}$ enforces all $\binom{N}{2}$ semi-local conditions simultaneously, in direct analogy with the 3-qubit construction of Eqs.~\eqref{eq:constants}. The global consistency condition
\begin{equation}
    N\, c_H^{(N)} + \tbinom{N}{2}\, c_V^{(N)} = \Lambda_N
    \label{eq:consistency}
\end{equation}
is automatically satisfied, since $\langle g | H_{\text{tot}}^{(N)} | g \rangle = 0$ follows from Eq.~\eqref{eq:zero_gs} and the decomposition $H_{\text{tot}}^{(N)} = \sum_n H_n + \sum_{i<j} V_{i,j}$.

\paragraph{Verification of QET protocol conditions.}
The Hamiltonian so constructed inherits all the structural properties required for a valid QET protocol:
\begin{enumerate}
    \item \textit{Zero ground-state energy}: $\langle g | H_{\text{tot}}^{(N)} | g \rangle = 0$ by Eq.~\eqref{eq:zero_gs}, and all subsystem expectations vanish by Eq.~\eqref{eq:zero_mean_all_N}.
    \item \textit{Non-commutativity}: $[H_n, V_{i,j}] = [h Z_n, k X_i X_j] \neq 0$ whenever $n \in \{i,j\}$, ensuring that the ground state is a non-trivially entangled state rather than a product state.
    \item \textit{Local passivity}: The constant shifts $c_H^{(N)}, c_V^{(N)}$ guarantee that $|g\rangle$ is locally passive, i.e., no local unitary operation alone can extract energy from the ground state~\cite{PuszWoronowicz, FreyFunoHotta}.
    \item \textit{Ground-state entanglement}: For any positive $h, k > 0$, the competition between the local $Z_n$ terms and the pairwise $X_i X_j$ interactions ensures that $|g\rangle$ is entangled across all subsystems, providing the essential quantum resource for energy teleportation.
\end{enumerate}

\paragraph{Computational limitations.}
While the above procedure is exact and constructive in principle, its practical realization faces significant challenges that grow with $N$:
\begin{enumerate}
    \item \textit{Exponential cost of spectral decomposition.} The determination of $\Lambda_N$ via Eq.~\eqref{eq:Lambda_fix} requires computing the lowest eigenvalue $E_0(h,k;0)$ of a $2^N \times 2^N$ matrix. For $N \ge 4$, full diagonalization becomes increasingly expensive, and the ground-state coefficients no longer collapse to a single parameter such as $M$ but proliferate in number and complexity.
    \item \textit{Null-space intractability.} Even after the global constant $\Lambda_N$ is determined, verifying the zero-mean-energy conditions~\eqref{eq:zero_mean_all_N} and decomposing the Hamiltonian into valid QET subsystem terms requires computing expectation values with respect to the ground state. For $N \ge 4$, the null space of the Hamiltonian restricted to specific symmetry sectors cannot always be obtained in closed form with current computational methods, rendering exact analytical expressions for $c_H^{(N)}$ and $c_V^{(N)}$ generally unavailable.
    \item \textit{Constraint proliferation.} The zero-mean-energy conditions constitute a system of $\binom{N}{2}+N+1$ equations (one for each $V_{i,j}$, each $H_n$, and the total Hamiltonian). For $N = 3$ we have shown that index symmetry reduces this to a pair of independent conditions; for larger $N$ with broken symmetry (e.g., non-uniform couplings), all constraints must be verified independently, and their number grows as $O(N^2)$.
\end{enumerate}
We emphasize, however, that the constructive procedure outlined above reduces the problem of building a valid $N$-qubit QET Hamiltonian to a single well-posed computational task: diagonalizing the bare Hamiltonian $H_{\text{tot}}^{(N)}(h,k;0)$ and extracting its lowest eigenvalue. For moderate qubit numbers ($N \lesssim 20$), this is readily achievable by exact numerical methods, while for larger systems, variational and tensor-network approaches (e.g., DMRG) provide efficient approximations of $E_0$ and $|g\rangle$. We therefore regard this framework as a practical pathway for systematic extension of QET protocols to higher-qubit systems, with the $N = 2$ and $N = 3$ constructions presented in this manuscript serving as analytically exact anchor points.

We further remark that the QET protocol itself is not specific to this particular ground state. As formulated by Hotta \cite{key18,key22}, the protocol requires only that the initial state be the entangled ground state of a Hamiltonian satisfying the zero-mean-energy condition \eqref{eq:zero_energy_all}. The state \eqref{eq:ground_state} is the ground state singled out by our 3-qubit Ising Hamiltonian, and the protocol parameters are optimized for it; in principle, however, any entangled ground state satisfying \eqref{eq:zero_energy_all} constitutes a valid resource for QET, with the achievable extraction efficiency determined by the entanglement structure of the particular state, as quantified by the energy-entropy bound of Section~\ref{sec:entropy_bound}.

\subsection{3-qubit QET - SIMO}

\begin{figure*}[!t]
\centering
\includegraphics[width=.82\textwidth]{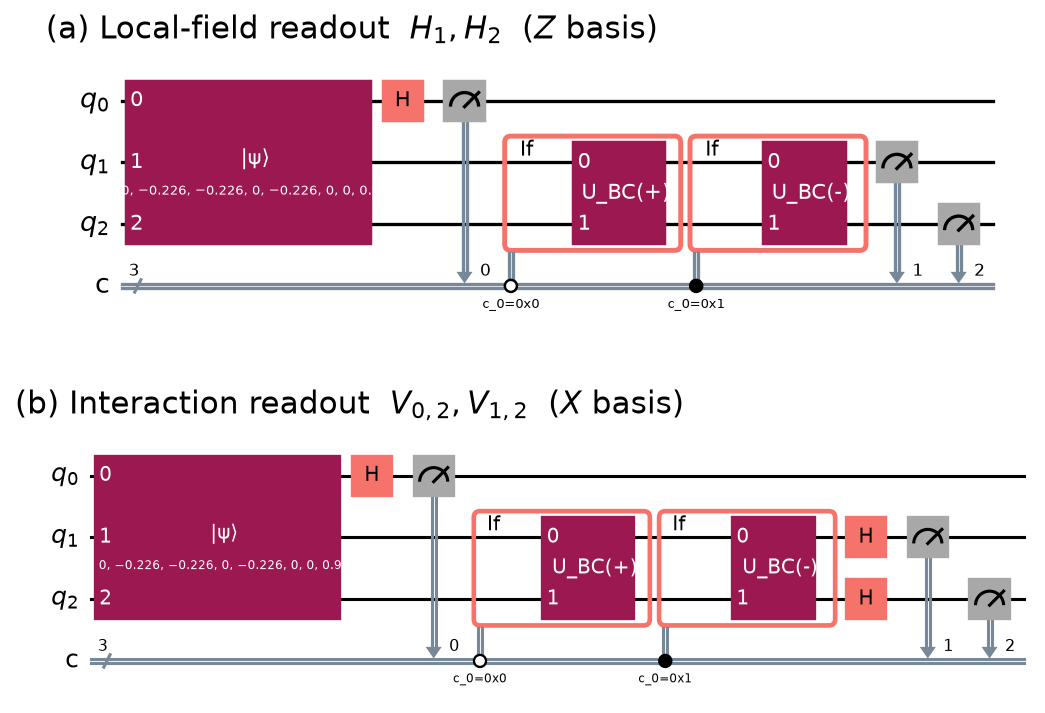}
\caption{SIMO with two receivers. Alice ($q_0$) measures $X_0$; conditioned on her outcome $\mu$, Bob ($q_2$) and Charlie ($q_1$) apply a joint two-qubit rotation $U_{BC}(\mu)$ that maps their conditional state to the ground state of the receiver Hamiltonian $H_\mu$, then read out their local fields $H_1,H_2$ (panel a, $Z$ basis) and interaction bonds $V_{0,1},V_{0,2},V_{1,2}$ (panel b, $X$ basis).}
\label{fig:simo_qc}
\end{figure*}

The SIMO (Single-Input Multiple-Output) configuration uses a single energy source, Alice, and two receivers, Bob (qubit 2) and Charlie (qubit 1), who jointly extract energy from their shared subsystem. Only Alice deposits energy, through a projective measurement of her transverse operator $X_0$; the two receivers then recover energy by a joint conditional operation, with no term of their region excluded from the accounting.

\subsubsection{Energy Deposit and Joint Extraction}

Once Alice observes $\mu \in \{-1, 1\}$, she communicates the outcome to both receivers within a time $t \ll 1/k$ (here $t = O(10)$ ns and $1/k = O(100)$ ns). Her projective measurement deposits $E_0 = c_H$ into the lattice and leaves her qubit in the pure eigenstate $|\mu\rangle$, fully disentangled from the receivers, $P_0(\mu)|g\rangle \propto |\mu\rangle \otimes |\psi_\mu\rangle$. Because $X_0$ commutes with every term of the receiver pair,
\begin{equation}
[X_0, H_1] = [X_0, H_2] = [X_0, V_{1,2}] = 0 ,
\label{eq:commute}
\end{equation}
the deposit injects no energy directly into the receivers; all subsequent dynamics are governed by the conditional Hamiltonian
\begin{equation}
\begin{aligned}
H_\mu &= \langle \mu | H_{\text{tot}} | \mu \rangle \\
&= h(Z_1+Z_2) + kX_1X_2 + \mu k(X_1+X_2) \\
&\quad + (2K+h-k)\,\mathbb{I},
\end{aligned}
\label{eq:cond_H_simo}
\end{equation}
a positive-semidefinite $4\times4$ operator on the receiver pair $(q_1,q_2)$. Acting jointly, Bob and Charlie can lower their energy no further than the ground-state value $\lambda_{\min}(H_\mu)$, so the extractable work is
\begin{equation}
W = E_0 - \sum_{\mu = \pm 1} p_\mu\, \lambda_{\min}(H_\mu),
\label{eq:w_simo}
\end{equation}
and the SIMO efficiency is $\eta_{\text{SIMO}} = W/E_0$. The optimal joint rotation $U_{BC}(\mu)$ maps $|\psi_\mu\rangle$ to the ground state of $H_\mu$; since $H_\mu$ and $|\psi_\mu\rangle$ are real, a real two-qubit rotation on $(q_1,q_2)$ attains the bound.

In contrast to the MISO entangler, the inter-receiver bond $V_{1,2}$ is fully retained in the accounting. At the joint optimum the two sender-receiver bonds are driven negative, $\langle V_{0,1}\rangle = \langle V_{0,2}\rangle < 0$, while the inter-receiver bond turns positive, $\langle V_{1,2}\rangle > 0$ (the energy the receivers spend on their mutual coupling in order to extract), and the on-site fields $\langle H_1\rangle = \langle H_2\rangle$ remain small. The work $W$ is therefore the honest net energy the two receivers extract, with no term excluded.

\subsubsection{Efficiency of the SIMO Protocol}

Table~\ref{tab:assist} lists the exact efficiency at the joint-receiver optimum. At $(h,k)=(1,3)$ and $(1,4)$ the deposit is $E_0 = 0.796$ and $0.770$, the extracted work is $W = 0.075$ and $0.077$, and the efficiency is $\eta_{\text{SIMO}} = W/E_0 \approx 9.4\%$ and $10.0\%$. Across the coupling range studied SIMO reaches $8\%$ to $10\%$, with the single sender Alice the only energy source. The efficiency depends only on the ratio $k/h$ and rises monotonically with it.

\begin{table}[!t]
\centering
\caption{Exact SIMO efficiency $\eta_{\text{SIMO}} = W/E_0$ at the joint-receiver optimum. $E_0 = c_H$ is Alice's deposit; $\langle V_{0,2}\rangle$ is a sender-receiver bond and $\langle V_{1,2}\rangle$ the inter-receiver bond (fully included); $W = E_0 - \sum_\mu p_\mu \lambda_{\min}(H_\mu)$ is the net extracted work.}
\label{tab:assist}
\begin{tabular}{|c|c|c|c|c|}
\hline
$(h,k)$ & $E_0$ & $\langle V_{0,2} \rangle$ & $\langle V_{1,2} \rangle$ & $\eta_{\text{SIMO}}$ \\ \hline
$(1,3)$ & $0.796$ & $-0.455$ & $+0.803$ & $9.4\%$ \\
$(1,4)$ & $0.770$ & $-0.614$ & $+1.138$ & $10.0\%$ \\ \hline
\end{tabular}
\end{table}

\subsubsection{Quantum Circuit Implementation for SIMO}

Figure~\ref{fig:simo_qc} shows the corresponding circuit. Alice measures $X_0$ (a Hadamard gate followed by a computational-basis measurement); conditioned on the outcome, Bob and Charlie apply the joint two-qubit rotation $U_{BC}(\mu)$ and read their local fields $H_1,H_2$ in the $Z$ basis and their bonds $V_{0,1},V_{0,2},V_{1,2}$ in the $X$ basis. Although the bonds are not local operators, the receivers obtain their expectation values through operations on their joint subsystem alone, conditioned on the received classical bit. By scanning the rotation we obtain $W > 0$ as theoretically predicted, confirming successful energy extraction.

\subsection{3-qubit QET - (MISO)}

\begin{figure*}[!t]
\centering
\includegraphics[width=.82\textwidth]{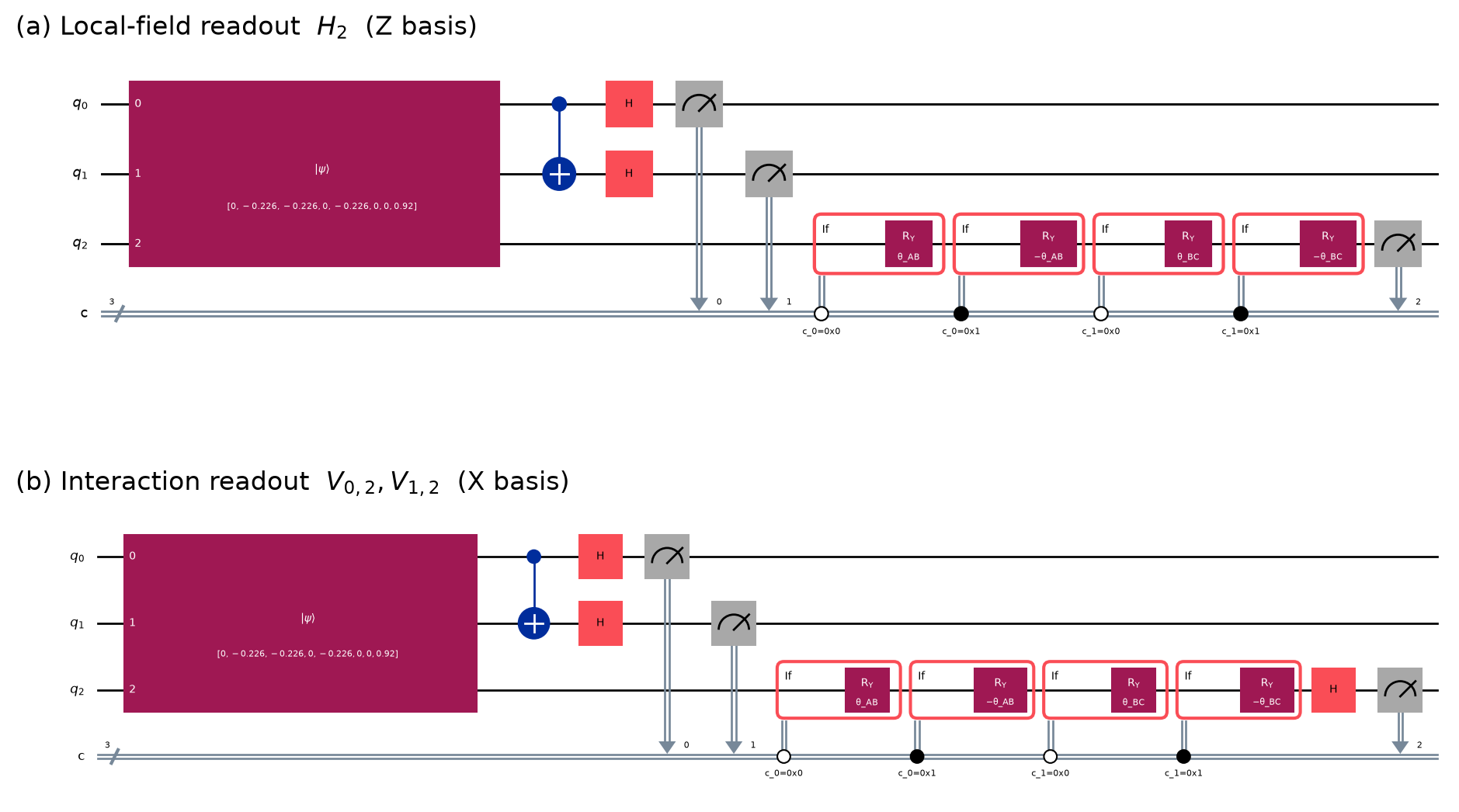}
\caption{MISO with a combined measurement. Alice ($q_0$) and Charlie ($q_1$) are entangled by a CNOT and then jointly measured in the $X$ basis; conditioned on the outcomes, Bob ($q_2$) applies $R_Y(\pm\theta_{AB})$ and $R_Y(\pm\theta_{BC})$ and reads out his local field $H_2$ (panel a, $Z$ basis) and interaction terms $V_{0,2},V_{1,2}$ (panel b, $X$ basis).}
\label{fig:miso_circuit}
\end{figure*}

In the MISO configuration, Alice (qubit 0) and Charlie (qubit 1) act as \emph{combined} senders, while Bob (qubit 2) is the single receiver. The two senders deposit energy through one joint measurement of their qubits, in contrast to the SIMO configuration, where a single sender acts. The protocol proceeds in two steps.

\subsubsection{Step 1: Combined Energy Deposit (Alice \& Charlie)}

The two senders first entangle their qubits with a controlled-NOT, $C \equiv \mathrm{CNOT}_{0\to1}$, and then measure both in the $X$ basis through the projectors $P_0(\mu_0)=\tfrac12(1+\mu_0 X_0)$ and $P_1(\mu_1)=\tfrac12(1+\mu_1 X_1)$, obtaining outcomes $\mu_0,\mu_1\in\{-1,1\}$. This single combined measurement deposits a joint energy into the ground state (whose energy is zero),
\begin{multline}
\langle E_{AC} \rangle = \sum_{\mu_0,\mu_1} \langle g | C^\dagger P_0(\mu_0) P_1(\mu_1)\, H_{\text{tot}} \\
\times\, P_1(\mu_1) P_0(\mu_0)\, C | g \rangle .
\label{eq:eac_deposit}
\end{multline}
Because the entangling step correlates the two senders, the deposit is a single combined input; for bookkeeping we attribute it symmetrically to the two senders, $E_a = E_c = \langle E_{AC}\rangle/2$, though it cannot be physically decomposed into independent Alice and Charlie measurements. This combined, two-sender deposit is what distinguishes MISO from the single-sender SIMO configuration. The entangler also transforms the sender-receiver bonds (for example $C^\dagger V_{0,2} C = k\,X_0X_1X_2 + v_c\,\mathbb{I}$), so that, unlike the separate-measurement case, the combined measurement injects energy into Bob's region, which Bob then extracts.

\subsubsection{Step 2: Energy Extraction (Bob)}

Once the combined measurement returns $\mu_0 \in \{-1,1\}$ and $\mu_1 \in \{-1,1\}$, the senders communicate the two outcomes to Bob via classical channels. Upon receiving this information, Bob applies a conditional unitary operation $U_b(\mu_0, \mu_1)$ to his qubit and subsequently measures his local energy. The classical communication must occur within a time $t$ that is significantly shorter than the coupling time scale, i.e., $t \ll 1/k$. In our experimental implementation, $t = O(10)$ ns while $1/k = O(100)$ ns, ensuring the validity of the protocol assumptions. The unitary operations $U_b(\mu_0)$ and $U_b(\mu_1)$ are given by:

\begin{equation}
U_b(\mu_{0}) = U_b(\mu_{1}) = \cos\phi \, I - i \mu \sin\phi \, Y_2 = R_Y(2\phi)
\end{equation}

The parameters $\xi$ and $\eta$ that determine the rotation angle $\phi$ are expressed as:

Since the three ground-state coefficients coincide ($M_1=M_2=M_3\equiv M$), these reduce to simple quadratics in the single parameter $M$:
\begin{align}
\xi  &= K - \tfrac{1}{2}k + kM + \left(2h - \tfrac{5}{2}k + 3K\right)M^{2}, \label{eq:xi_miso}\\
\eta &= -k + 2(h+k)\,M + (3k - 2h)\,M^{2}, \label{eq:eta_miso}
\end{align}
with $K=\sqrt{h^2+hk+k^2}$ and $M=k/(2h+k+2K)$, so that $\xi$ and $\eta$ depend only on $h$ and $k$.

The angle $\phi$ is then determined by:

\begin{equation}
\cos(2\phi) = \frac{\xi}{\sqrt{\xi^2 + \eta^2}},
\quad
\sin(2\phi) = \frac{\eta}{\sqrt{\xi^2 + \eta^2}}
\label{eq:phi_miso}
\end{equation}

For the 3-qubit QET system, the values of $\xi$ and $\eta$ can be generalized using the following quantum mechanical expectation values:

\begin{equation}
\begin{aligned}
\xi &= \langle g| \sigma_B H \sigma_B |g\rangle, \\
\eta &= \eta_A = \eta_C = \langle g| \sigma_A \dot{\sigma}_B |g\rangle = \langle g| \sigma_C \dot{\sigma}_B |g\rangle ,
\end{aligned}
\end{equation}

where $\sigma_B$ represents the Pauli operation applied by Bob, and the time derivative operator $\dot{\sigma}_B$ is defined as:
\begin{equation}
\dot{\sigma}_B = i [H_b, \sigma_B] = [H, \sigma_B]
\end{equation}

The average state after the combined measurement and Bob's conditional rotation $U_b(\mu_0,\mu_1)$ is:

\begin{multline}
\rho_{\text{QET}} = \sum_{\mu_0, \mu_1} U_b(\mu_0,\mu_1)\, P_1(\mu_1) P_0(\mu_0)\, C \\
\times\, |g\rangle\langle g|\, C^\dagger\, P_0(\mu_0) P_1(\mu_1)\, U_b^\dagger(\mu_0,\mu_1).
\end{multline}

The work extracted by Bob is the decrease of the total energy relative to the combined deposit,
\begin{equation}
W = \langle E_{AC} \rangle - \text{Tr} [ \rho_{\text{QET}} H_{\text{tot}} ],
\label{eq:e2_miso}
\end{equation}
Bob's conditional rotation $U_b(\mu_0,\mu_1)$ acts only on qubit 2, so it leaves the senders' deposit $\langle E_S\rangle = \langle H_0+H_1+V_{0,1}\rangle$ untouched and changes only Bob's region $H_2+V_{0,2}+V_{1,2}$. The entangling step of the combined measurement, however, already places part of the deposit directly into Bob's bond, $\langle E_b^{(0)}\rangle = \langle H_2+V_{0,2}+V_{1,2}\rangle$ evaluated right after the measurement, so that $\langle E_{AC}\rangle = \langle E_S\rangle + \langle E_b^{(0)}\rangle$. Because the induced map is \textit{not} unitary, Bob's rotation then drives his region below its ground-state value, $\langle E_b\rangle = \text{Tr}[\rho_{\text{QET}}(H_2+V_{0,2}+V_{1,2})] < 0$ with $V_{0,2}=V_{1,2}<0$, the negative-energy signature of QET [Eq.~\eqref{eq:eb_miso}]. Since the entangler-deposited part $\langle E_b^{(0)}\rangle$ is wired directly into Bob's bond rather than teleported, we report the \emph{net} efficiency: the genuinely teleported negative energy normalized by the energy deposited at the senders,
\begin{equation}
\eta_{\text{MISO}} = \frac{-\langle E_b\rangle}{\langle E_S\rangle} = \frac{-\langle E_b\rangle}{\langle E_{AC}\rangle - \langle E_b^{(0)}\rangle},
\label{eq:eta_miso_def}
\end{equation}
with energy conservation respected, $0 < -\langle E_b\rangle \le \langle E_S\rangle$.

\subsubsection{Physical Interpretation and Measurement Strategy for MISO}

The MISO (Multiple-Input Single-Output) model represents a configuration where two senders (Alice and Charlie) deposit energy through a single combined (entangle-then-measure) measurement, and a single receiver (Bob) extracts the energy through conditional operations. Similar to the minimal QET model, the interaction Hamiltonian $V$ and local Hamiltonian $H_2$ do not commute, necessitating separate measurements of these observables. Specifically, Bob performs independent measurements of $V_{0,2}$, $V_{1,2}$, and $H_2$, obtaining the expectation values $\langle V_{0,2} \rangle$, $\langle V_{1,2} \rangle$, and $\langle H_2 \rangle$ through statistical averaging over multiple experimental runs.

Our theoretical and experimental analysis reveals that $\langle V_{0,2} \rangle$ and $\langle V_{1,2} \rangle$ are consistently negative, while $\langle H_2 \rangle$ is positive. Bob therefore reads all three terms: the two negative interaction bonds are the thermodynamic signature of the teleported energy and its dominant contribution, partially offset by the positive on-site field $\langle H_2 \rangle$. Together they fix Bob's region energy $\langle E_b\rangle$ [Eq.~\eqref{eq:eb_miso}] and hence the conserved work $W$ [Eq.~\eqref{eq:e2_miso}].

\subsubsection{Quantum Circuit Implementation for MISO}

After the combined measurement, the interaction terms read by Bob are
\begin{multline*}
\langle V_{0,2}(\mu_0,\mu_1)\rangle = \langle g | C^\dagger P_0(\mu_0) P_1(\mu_1) U_b^\dagger(\mu_0,\mu_1) \\
\times V_{0,2}\, U_b(\mu_0,\mu_1) P_1(\mu_1) P_0(\mu_0) C | g \rangle ,
\end{multline*}
\begin{multline*}
\langle V_{1,2}(\mu_0,\mu_1)\rangle = \langle g | C^\dagger P_0(\mu_0) P_1(\mu_1) U_b^\dagger(\mu_0,\mu_1) \\
\times V_{1,2}\, U_b(\mu_0,\mu_1) P_1(\mu_1) P_0(\mu_0) C | g \rangle .
\end{multline*}

Figure \ref{fig:miso_circuit} shows the corresponding circuit: the CNOT entangler on the sender pair, their joint $X$-basis measurement, and Bob's conditional rotations, with $V_{0,2},V_{1,2}$ read in the $X$ basis (panel b) and $H_2$ in the $Z$ basis (panel a). Although $V$ is not a local operator, Bob obtains its expectation value through local operations alone, conditioned on the classical outcomes.

The average energy of Bob's region after extraction is
\begin{multline}
\langle E_b \rangle = \sum_{\mu_0, \mu_1} \langle g | C^\dagger P_0(\mu_0) P_1(\mu_1) U_b^\dagger(\mu_0,\mu_1) \\
\times (H_2 + V_{0,2} + V_{1,2})\, U_b(\mu_0,\mu_1) \\
\times P_1(\mu_1) P_0(\mu_0) C |g\rangle .
\label{eq:eb_miso}
\end{multline}

Optimising Bob's conditional rotation for each outcome pair drives $\langle V_{0,2}\rangle$ and $\langle V_{1,2}\rangle$ negative. At $(h,k)=(1,3)$ and $(1,4)$ the combined deposit is $\langle E_{AC}\rangle = 3.47$ and $4.11$, of which $\langle E_b^{(0)}\rangle = 0.94$ and $1.29$ is placed directly in Bob's bond, leaving the sender deposit $\langle E_S\rangle = 2.53$ and $2.83$. Bob's rotation then drives his region to $\langle E_b\rangle = -0.86$ and $-1.19$, giving a net efficiency $\eta_{\text{MISO}} = -\langle E_b\rangle/\langle E_S\rangle \approx 34\%\text{ to }42\%$, well above the minimal 2-qubit value, with energy conservation respected ($-\langle E_b\rangle \le \langle E_S\rangle$).

\begin{figure*}[!t]
\centering
\begin{subfigure}[b]{0.49\textwidth}
\includegraphics[width=\textwidth]{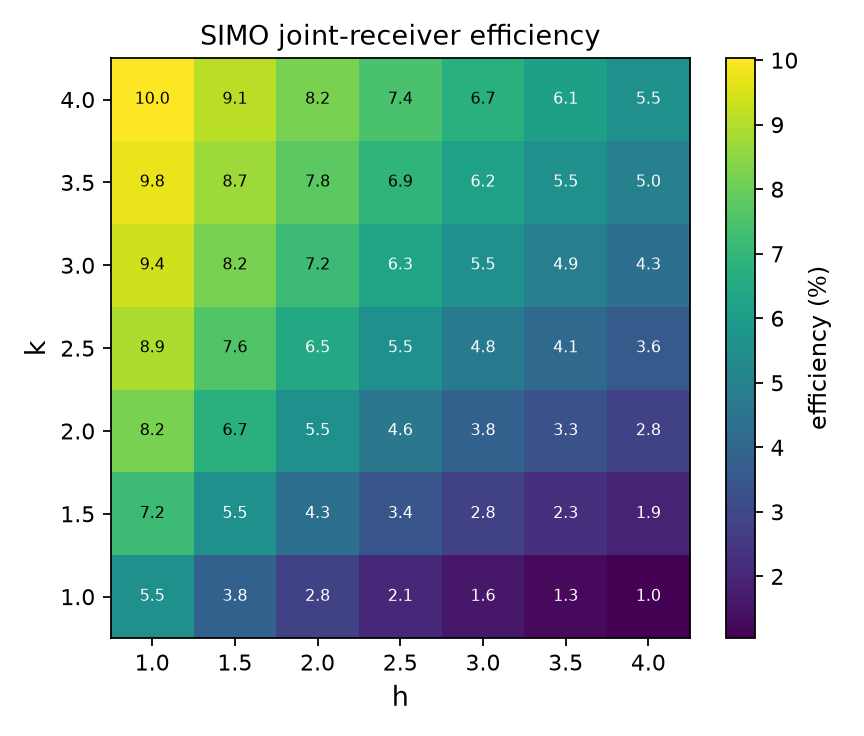}
\caption{SIMO}
\end{subfigure}\hfill
\begin{subfigure}[b]{0.49\textwidth}
\includegraphics[width=\textwidth]{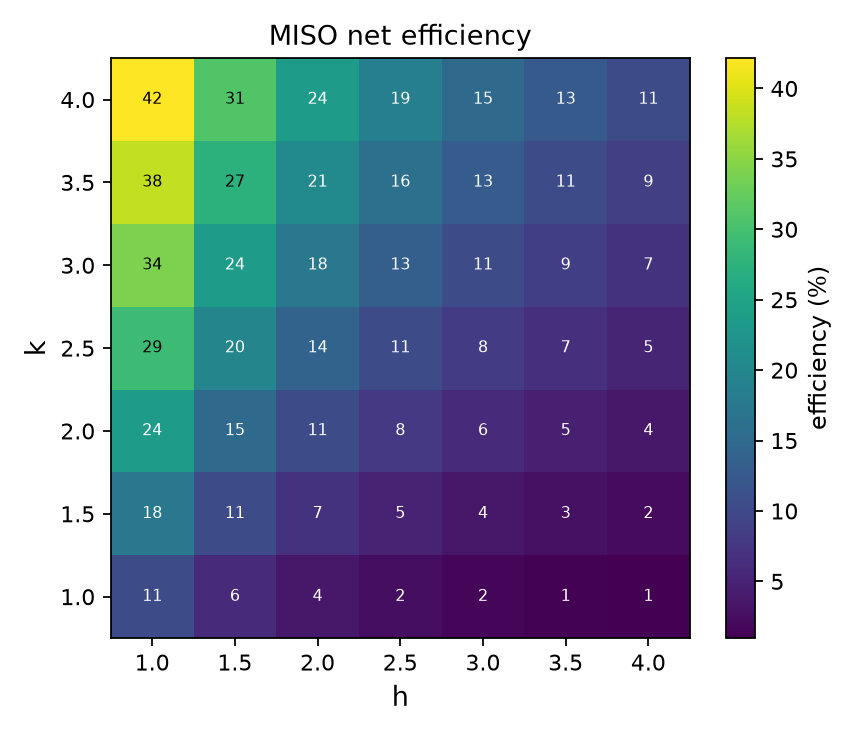}
\caption{MISO}
\end{subfigure}
\caption{Exact efficiency over the $(h,k)$ plane ($h,k$ from $1$ to $4$ in steps of $0.5$). (a) SIMO efficiency $\eta_{\text{SIMO}}=W/E_0$ for the joint two-receiver extraction. (b) MISO net efficiency $\eta_{\text{MISO}}=-\langle E_b\rangle/\langle E_S\rangle$ for the combined (entangle-then-measure) deposit. Both rise monotonically with the coupling ratio $k/h$ and depend only on $k/h$ by scale invariance.}
\label{fig:heatmaps}
\end{figure*}

Figure~\ref{fig:heatmaps} maps the exact efficiency of both protocols over the $(h,k)$ plane; both increase monotonically with the coupling ratio $k/h$, with the SIMO efficiency reaching $8\%$ to $10\%$ and the MISO net efficiency $34\%$ to $42\%$ in the regime $k/h=3$ to $4$ used in our experiments.

\section{Quantum Gates and Measurement Framework}

In this section, we provide a comprehensive explanation of the fundamental quantum gates and measurement procedures utilized in our 3-qubit extended QET protocol. These form the building blocks for constructing the ground state preparation circuits and implementing the energy teleportation protocol.

\subsection{Single-Qubit Pauli Operators and Basis States}

We employ the standard single-qubit Pauli operators and Hadamard gate, whose matrix representations in the computational basis are:

\begin{align}
    X &= \begin{pmatrix}
    0 & 1 \\
    1 & 0
    \end{pmatrix}, \quad
    Y = \begin{pmatrix}
    0 & -i \\
    i & 0
    \end{pmatrix}, \\
    Z &= \begin{pmatrix}
    1 & 0 \\
    0 & -1
    \end{pmatrix}, \quad
    H = \frac{1}{\sqrt{2}} \begin{pmatrix}
    1 & 1 \\
    1 & -1
    \end{pmatrix}.
\end{align}

For our 3-qubit system, these operators act on individual qubits and are denoted by subscripts indicating the target qubit. For example, $X_0$ denotes the Pauli $X$ operator acting on qubit 0, with the identity operator acting on qubits 1 and 2, i.e., $X_0 = X \otimes I \otimes I$.

The computational basis states for a single qubit are:
\begin{equation}
|0\rangle = \begin{pmatrix} 1 \\ 0 \end{pmatrix}, \quad |1\rangle = \begin{pmatrix} 0 \\ 1 \end{pmatrix}
\end{equation}

These are the eigenstates of the Pauli $Z$ operator with eigenvalues $+1$ and $-1$, respectively:
\begin{equation}
Z|0\rangle = |0\rangle, \quad Z|1\rangle = -|1\rangle
\end{equation}

For the 3-qubit system, the computational basis consists of eight states: $|000\rangle, |001\rangle, |010\rangle, |011\rangle, |100\rangle, |101\rangle, |110\rangle, |111\rangle$, which form a complete orthonormal basis for the $2^3 = 8$-dimensional Hilbert space.

\subsection{Hadamard Basis and Eigenstates of Pauli X}

In addition to the computational basis, we utilize the Hadamard basis, defined by the eigenstates of the Pauli $X$ operator:
\begin{equation}
|\pm\rangle = \frac{1}{\sqrt{2}}(|0\rangle \pm |1\rangle)
\end{equation}

These states satisfy the eigenvalue equations:
\begin{equation}
X|+\rangle = |+\rangle, \quad X|-\rangle = -|-\rangle
\end{equation}

The transformation between the computational and Hadamard bases is accomplished by applying the Hadamard gate:
\begin{equation}
H|0\rangle = |+\rangle = \frac{1}{\sqrt{2}}(|0\rangle + |1\rangle), \quad H|1\rangle = |-\rangle = \frac{1}{\sqrt{2}}(|0\rangle - |1\rangle)
\end{equation}

In the QET protocol, Alice (qubit 0) and Charlie (qubit 1) perform projective measurements in the Hadamard basis by measuring their local Pauli $X$ operators. By observing the eigenvalues $\mu \in \{+1, -1\}$ of their respective local $X$ operators, Alice and Charlie obtain measurement outcomes that are subsequently communicated to Bob via classical channels. This classical information enables Bob to perform the appropriate conditional unitary operations on his qubit (qubit 2).

\subsection{Single-Qubit Rotation Gates}

The single-qubit rotation gates about the $X$, $Y$, and $Z$ axes are fundamental operations in our protocol, particularly for the conditional operations applied by the receivers. These rotations are defined as:

\begin{equation}
    R_X(\phi) = e^{-i\frac{\phi}{2}X} = \begin{pmatrix}
    \cos\frac{\phi}{2} & -i\sin\frac{\phi}{2} \\
    -i\sin\frac{\phi}{2} & \cos\frac{\phi}{2}
    \end{pmatrix}
\end{equation}

\begin{equation}
    R_Y(\phi) = e^{-i\frac{\phi}{2}Y} = \begin{pmatrix}
    \cos\frac{\phi}{2} & -\sin\frac{\phi}{2} \\
    \sin\frac{\phi}{2} & \cos\frac{\phi}{2}
    \end{pmatrix}
\end{equation}

\begin{equation}
    R_Z(\phi) = e^{-i\frac{\phi}{2}Z} = \begin{pmatrix}
    e^{-i\frac{\phi}{2}} & 0 \\
    0 & e^{i\frac{\phi}{2}}
    \end{pmatrix}
\end{equation}

In our 3-qubit extended QET protocol, the $R_Y$ rotation gates play a crucial role in both ground state preparation and in the conditional operations performed by the receiver Bob. The rotation angle $\phi$ is determined by the parameters $\xi$ and $\eta$ as specified in Eqs. \eqref{eq:xi_miso}, \eqref{eq:eta_miso}, and \eqref{eq:phi_miso}, which depend on the system parameters $h$ and $k$, as well as the ground state properties encoded in the coefficient $M$.

For the 3-qubit system, when a rotation gate acts on a specific qubit $i \in \{0, 1, 2\}$, it is implicitly understood that the operation is $R_Y(\phi)_i = I \otimes \cdots \otimes R_Y(\phi) \otimes \cdots \otimes I$, where the rotation acts on qubit $i$ and the identity operator acts on all other qubits. This notation is used throughout our quantum circuit implementations for both the MISO and SIMO configurations of the extended QET protocol.

 \section{Ground-State Preparation and Measurement Procedures}

In this section, we describe the controlled gate operations used to prepare the ground state $|g\rangle$ and the measurement procedures employed to estimate the local and interaction energies. The gate operations for the 3-qubit system build upon the two-qubit operations used in the minimal QET model, extended to accommodate the additional qubit.

\subsection{Controlled Operations and Ground State Preparation}

In general, a controlled $U$ operation $\Lambda(U)$ is defined by:
\begin{equation}
    \Lambda(U) = |0\rangle\langle 0| \otimes I + |1\rangle\langle 1| \otimes U 
\end{equation}

Following the well-known relation for CNOT gates, $\text{CNOT}(a|0\rangle + b|1\rangle)|0\rangle = a|00\rangle + b|11\rangle$, we construct the ground state for our extended QET protocol through a sequence of controlled operations and rotation gates. The ground state $|g\rangle$ is prepared by applying the following sequence of operations to the initial state $|000\rangle$:

\begin{equation}
\scalebox{0.75}{$
\begin{aligned}
|g\rangle = \Big[ &\left(I \otimes \left( |0\rangle\langle 0| \otimes I + |1\rangle\langle 1| \otimes X \right) \right) 
\cdot \left(I \otimes R_y(\theta) \otimes I\right) \cdot \left(X \otimes I \otimes I\right) \\
&\cdot \left( |0\rangle\langle 0| \otimes I \otimes I + |1\rangle\langle 1| \otimes I \otimes X \right) 
\cdot \left( |0\rangle\langle 0| \otimes I \otimes I + |1\rangle\langle 1| \otimes X \otimes I \right) \\
&\cdot \left( R_y(\theta) \otimes I \otimes I \right) \cdot \left(X \otimes I \otimes I \right) \Big] |000\rangle
\end{aligned}
$}
\end{equation}

This operation generates a superposition of basis states with specific coefficients:
\begin{equation}
|g\rangle = a \ket{001} + b \ket{010} + c \ket{100} + d \ket{111}
\end{equation}

where the coefficients $a$, $b$, $c$, and $d$ are determined by the rotation angle $\theta$ and are related to the parameter $M$ defined in the Hamiltonian construction. Specifically, these coefficients ensure that $|g\rangle$ satisfies the zero mean energy condition for the total Hamiltonian $H_{\text{tot}}$.

\subsection{Measurement of Local Observables}

The measurement outcomes are obtained as bit strings $b_0 b_1 b_2 \in \{000, 001, 010, 011, 100, 101, 110, 111\}$, where each bit corresponds to the measurement result of a single qubit. Since the eigenvalues of the Pauli $Z$ operator are $\pm 1$, we convert bit values to eigenvalues using the transformation $1 - 2b_i$, where $b_i \in \{0, 1\}$ maps to $\{+1, -1\}$.

Let $n_{\text{shots}}$ denote the total number of times the quantum circuit is executed, and let $\text{count}_{b_0 b_1 b_2}$ represent the number of times the specific bit string $b_0 b_1 b_2$ is observed. The probability of obtaining a particular bit string is then $P_{b_0 b_1 b_2} = \text{count}_{b_0 b_1 b_2} / n_{\text{shots}}$. The expectation value of the local observable $Z_i$ for qubit $i$ is computed as:

\begin{equation}
    \langle Z_i \rangle = \sum_{b_0, b_1, b_2} (1 - 2b_i) \frac{\text{count}_{b_0 b_1 b_2}}{n_{\text{shots}}}
\end{equation}

\subsection{Measurement of Two-Qubit Interaction Terms}

To measure the interaction Hamiltonian terms $X_i X_j$, we apply Hadamard gates to both qubits $i$ and $j$ before performing computational basis measurements. The Hadamard transformation maps the computational basis states $|0\rangle$ and $|1\rangle$ to the eigenstates of the Pauli $X$ operator, $|+\rangle = \frac{1}{\sqrt{2}}(|0\rangle + |1\rangle)$ and $|-\rangle = \frac{1}{\sqrt{2}}(|0\rangle - |1\rangle)$, with eigenvalues $+1$ and $-1$, respectively.

Following the Hadamard transformations, we obtain measurement outcomes as bit strings $b_0 b_1 b_2 \in \{000, 001, 010, 011, 100, 101, 110, 111\}$. These bit strings are converted to the eigenvalues of $X_i X_j$ using the product $(1 - 2b_i)(1 - 2b_j)$. The expectation value of the two-qubit interaction observable $X_i X_j$ is then computed as:

\begin{equation}
    \langle X_i X_j \rangle = \sum_{b_0, b_1, b_2} (1 - 2b_i)(1 - 2b_j) \frac{\text{count}_{b_0 b_1 b_2}}{n_{\text{shots}}}
\end{equation}

This measurement procedure allows us to extract the expectation values of both local Hamiltonians $H_n$ (via $Z_i$ measurements) and interaction Hamiltonians $V_{i,j}$ (via $X_i X_j$ measurements), which are essential for implementing and verifying the extended QET protocol on quantum hardware.

\section{Extended QET on IBM Quantum Environment}

In this section, we describe the implementation of the extended QET protocol on real quantum hardware, with particular emphasis on executing conditional operations that are not natively supported by most contemporary quantum computing platforms. As illustrated in the QET protocol, Bob's (and Charlie's in the extended model) operations must be selected based on the measurement outcomes obtained by Alice, creating a need for mid-circuit conditional logic.

\subsection{Deferred Measurement and Conditional Operations}

QET can be implemented seamlessly on quantum hardware that does not support mid-circuit conditional statements by utilizing the deferred measurement principle. By postponing Alice's measurement until the conclusion of the quantum circuit, we obtain results that are mathematically equivalent to those obtained with immediate measurement and conditional operations. The conditional operations can be constructed using a controlled $U$ gate, defined as $\Lambda(U) = |0\rangle\langle 0| \otimes I + |1\rangle\langle 1| \otimes U$, and an anti-controlled $U$ gate $(X \otimes I)\Lambda(U)(X \otimes I)$. Figure~\ref{fig:example} shows the direct conditional implementation, with Bob's outcome-conditioned rotations highlighted in the \textsf{If} frames; replacing these conditional operations by the deferred-measurement construction described above yields a mathematically equivalent circuit.

\subsection{Experimental Setup and Quantum Hardware}

\begin{figure}[!t]
    \centering
    \includegraphics[width=\columnwidth]{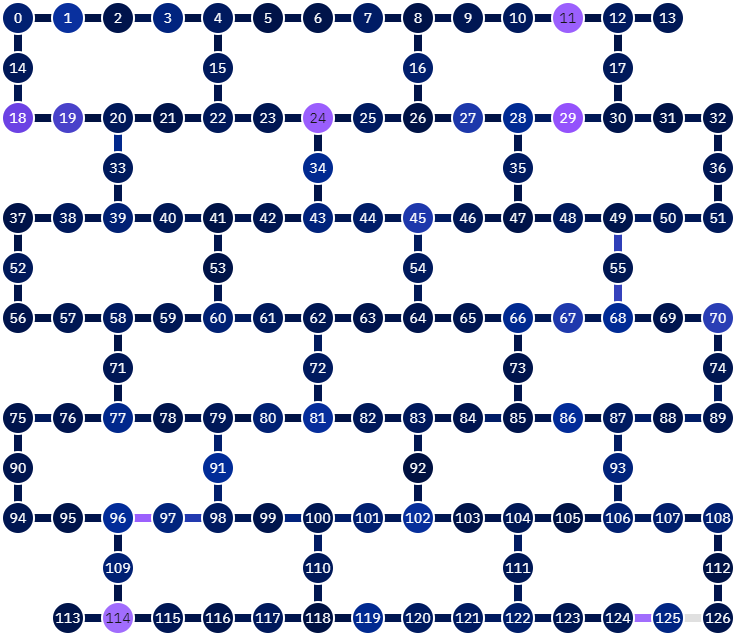}\\
    \includegraphics[width=\columnwidth]{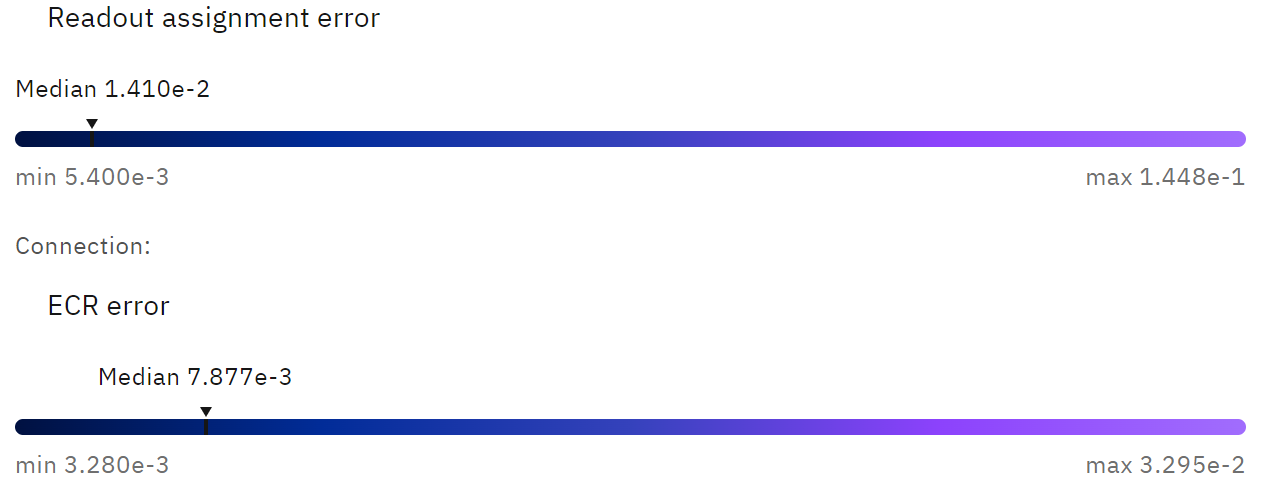}
    \caption{Qubit Map ibm\_brisbane}
    \label{fig:brisbane_map}
\end{figure}

\begin{figure}[!t]
    \centering
    \includegraphics[width=\columnwidth]{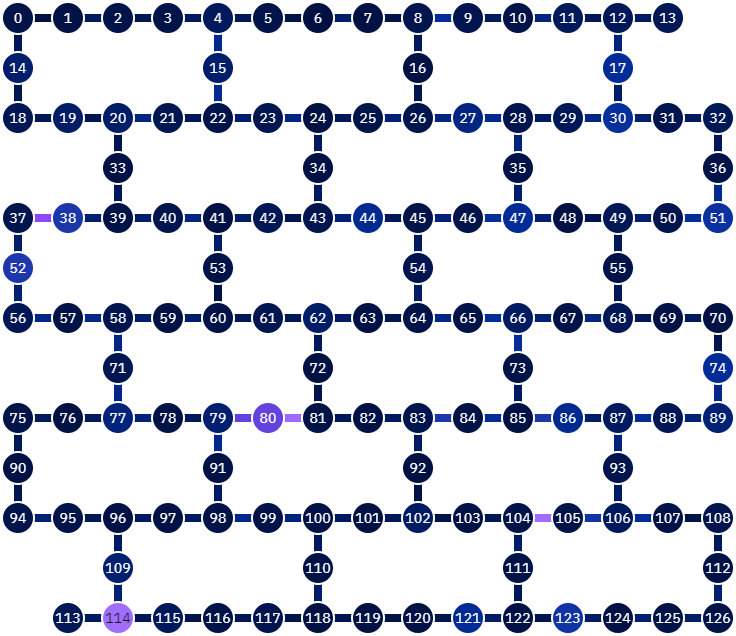}\\
    \includegraphics[width=\columnwidth]{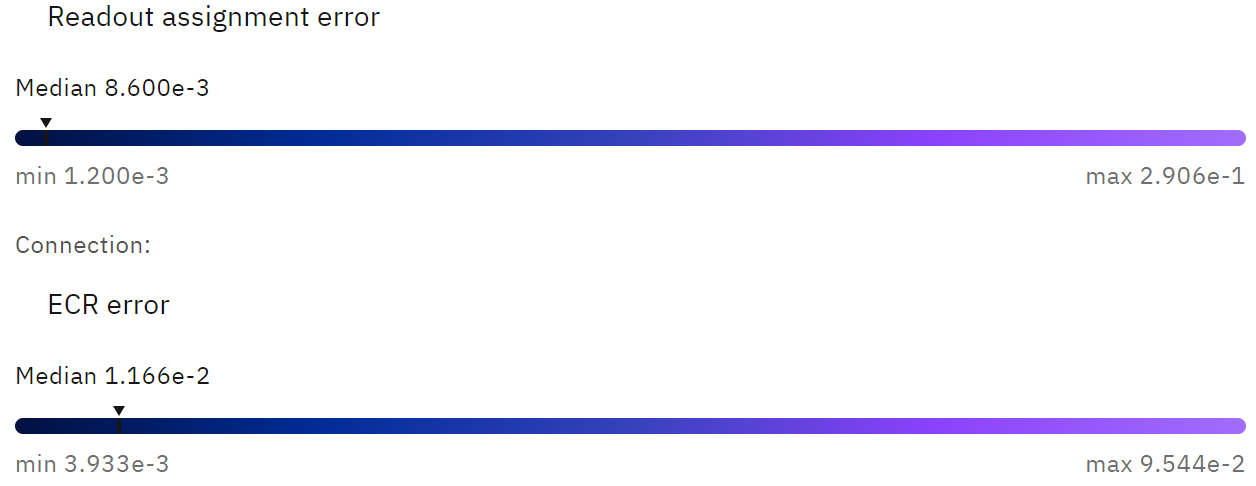}
    \caption{Qubit Map ibm\_kyiv}
    \label{fig:kyiv_map}
\end{figure}

\begin{figure}[!t]
    \centering
    \includegraphics[width=\columnwidth]{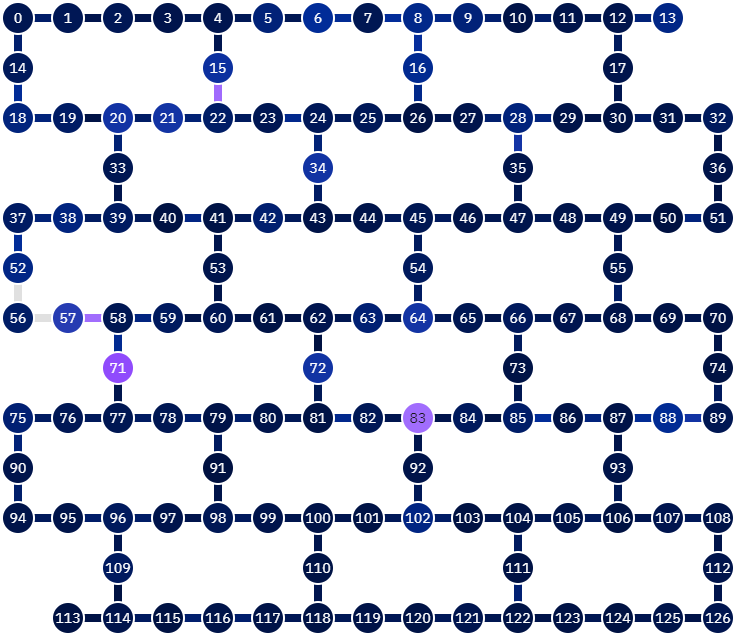}\\
    \includegraphics[width=\columnwidth]{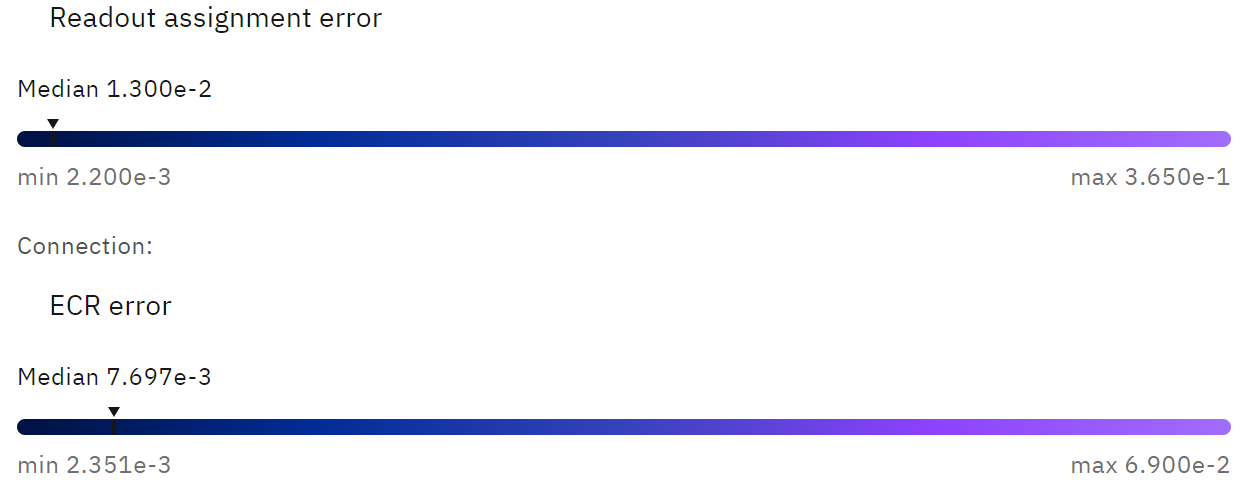}
    \caption{Qubit Map ibm\_sherbrooke}
    \label{fig:sherbrooke_map}
\end{figure}

For the implementation of our extended QET model, we conducted quantum computations on three distinct IBM quantum hardware platforms: \textit{ibm\_brisbane}, \textit{ibm\_sherbrooke}, and \textit{ibm\_kyiv}. The detailed characteristics and topological properties of each quantum computer are presented in Figs.~\ref{fig:brisbane_map}, \ref{fig:kyiv_map}, and \ref{fig:sherbrooke_map}, respectively, with the corresponding qubit-level calibration data summarized in Table~\ref{tab:calibration_data}. These devices feature different qubit connectivity graphs, where direct CNOT gates can only be applied to qubits connected by an edge in the hardware topology. For our experiments, we strategically selected connected qubit pairs with relatively low gate errors to maximize the fidelity of our implementations.

We executed the extended QET protocol on three 127-qubit IBM Quantum Eagle processors—\texttt{ibm\_brisbane}, \texttt{ibm\_sherbrooke}, and \texttt{ibm\_kyiv}—which feature fixed-frequency transmon qubits arranged in a ``Heavy Hex'' lattice topology. The device structures (Figs.~\ref{fig:brisbane_map} to \ref{fig:sherbrooke_map}) illustrate the specific connectivity graphs where direct two-qubit gates are restricted to physically coupled neighbors, with color gradients visualizing calibration data such as readout and ECR error rates. While \texttt{ibm\_sherbrooke} demonstrated the highest gate fidelity with a best-case ECR error of $2.35 \times 10^{-3}$, all systems exhibited robust coherence with median $T_1$ times exceeding $220~\mu\text{s}$, allowing us to validate our results against \texttt{qasm\_simulator} benchmarks by mapping circuits to high-fidelity qubit chains.

The experimental measurements across all three quantum devices exhibited a high degree of consistency, demonstrating the robustness of the extended QET protocol across different hardware architectures. Additionally, we performed comprehensive simulations using the \textit{qasm\_simulator}, which provides classical emulation of quantum gate operations on circuits identical to those executed on physical quantum hardware. The simulator results serve as a benchmark for evaluating the performance of real quantum devices.

\begin{figure}[!t]
    \centering
    \begin{subfigure}[b]{\linewidth}
        \centering
        \includegraphics[width=.9\linewidth]{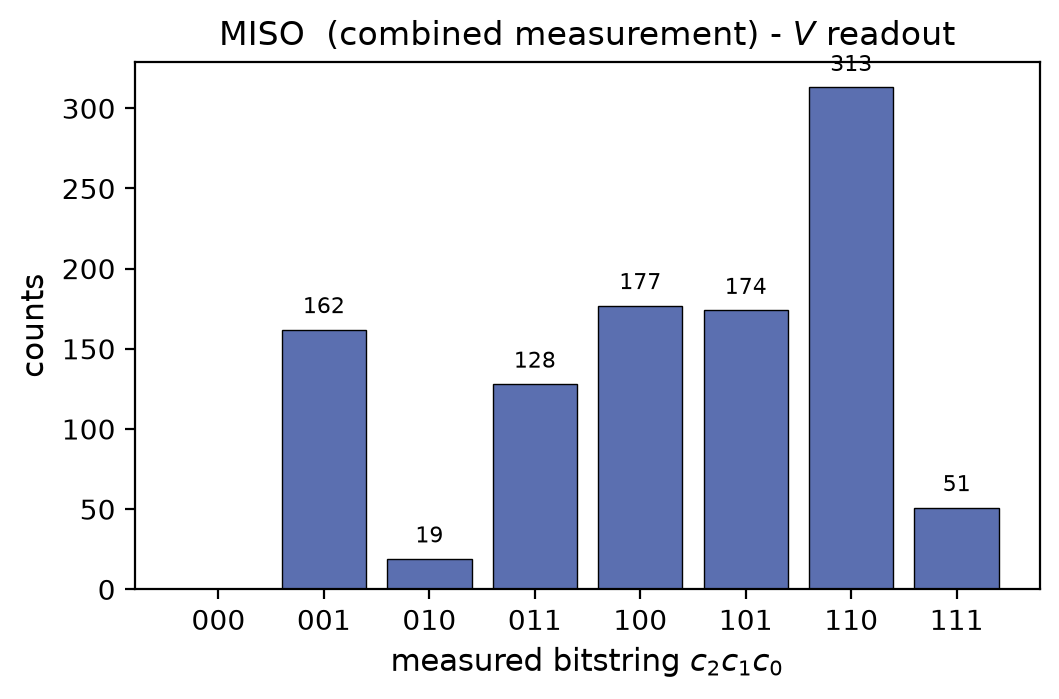}
        \caption{Bob Measures $V_{1,2}$ \& $V_{0,2}$}
        \label{fig:example1}
    \end{subfigure}\\
    \begin{subfigure}[b]{\linewidth}
        \centering
        \includegraphics[width=.9\linewidth]{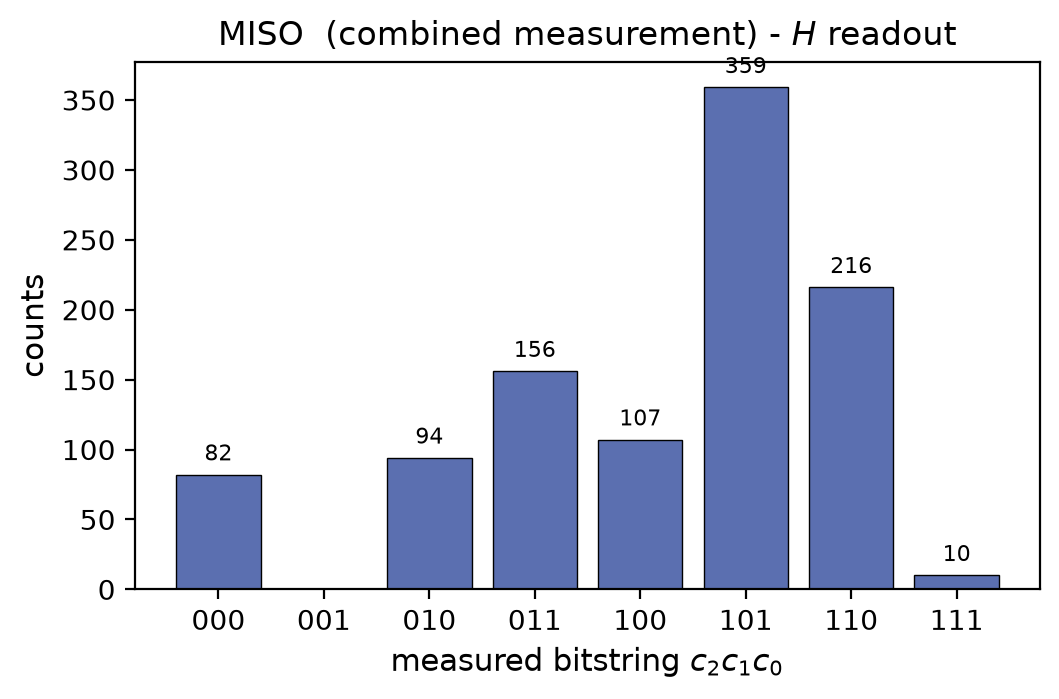}
        \caption{Bob Measures $H_{2}$}
        \label{fig:example2}
    \end{subfigure}
    \caption{Enhanced QET (MISO). ( Total Shots: 1024 )}
    \label{fig:miso_results}
\end{figure}

\begin{figure}[!t]
    \centering
    \begin{subfigure}[b]{\linewidth}
        \centering
        \includegraphics[width=.9\linewidth]{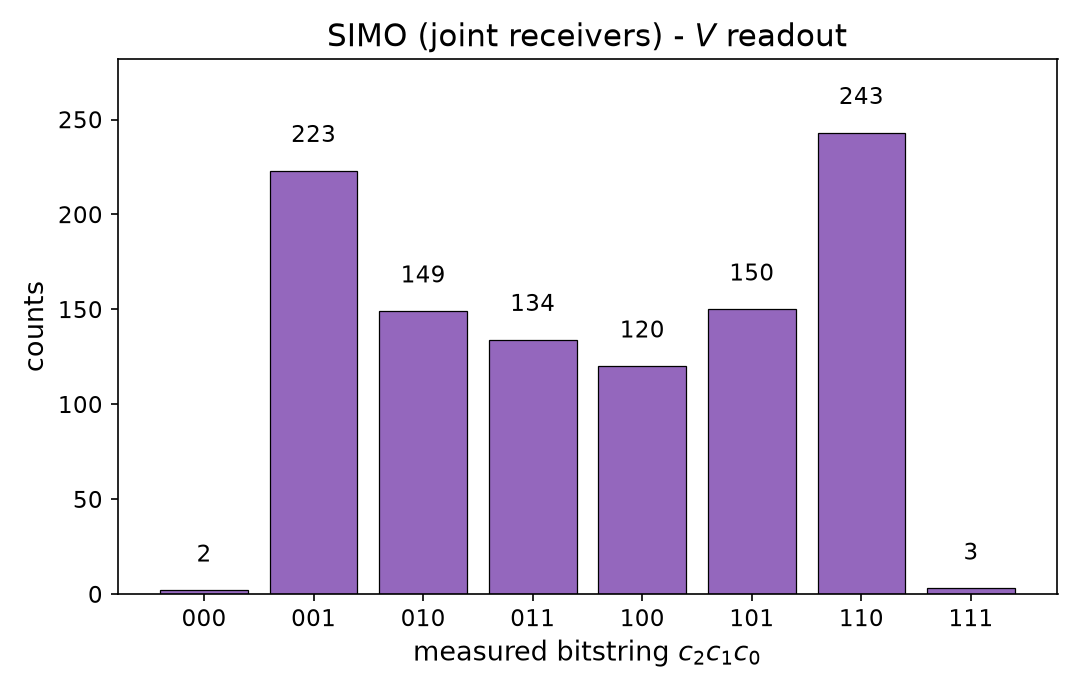}
        \caption{Receivers read the interaction bonds $V_{0,1},V_{0,2},V_{1,2}$ ($X$ basis)}
        \label{fig:example11}
    \end{subfigure}\\
    \begin{subfigure}[b]{\linewidth}
        \centering
         \fbox{\includegraphics[width=.9\linewidth]{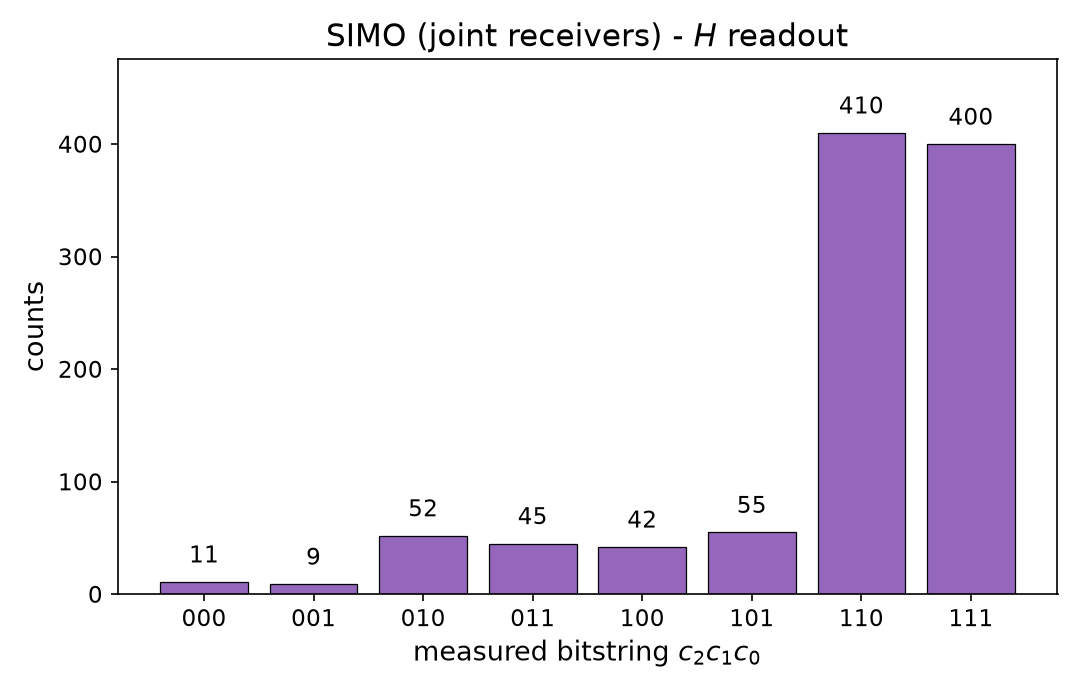}}
        \caption{Receivers read the local fields $H_1,H_2$ ($Z$ basis)}
        \label{fig:example22}
    \end{subfigure}
    \caption{Enhanced QET (SIMO). ( Total Shots: 1024 )}
    \label{fig:simo_results}
\end{figure}

\subsection{Experimental Results and Error Mitigation}

Our experimental implementation successfully demonstrated negative energy extraction, with $\langle E_b \rangle < 0$, validating the theoretical predictions of the extended QET protocol. Remarkably, the magnitude of $\langle V_b \rangle$ obtained in our 3-qubit system surpasses the results achievable with the 2-qubit minimal QET model. The measurement closest to the exact analytical value was $V_b = -0.198$ for parameters $(h = 1, k = 4)$ on \textit{ibm\_sherbrooke}, achieving approximately 91\% accuracy relative to the theoretical prediction.

As emphasized in Hotta's seminal works \cite{key5,key6,key7,key8,key9,key10,key11,key22}, it is fundamentally impossible for any unitary operation to produce $\langle E_b \rangle < 0$ after the sender's measurement of $X_i$ (where $i \in \{0, 1\}$), as shown in Eqs. \eqref{eq:eb_miso} and \eqref{eq:w_simo}. The successful observation of negative energy extraction requires precise classical communication between senders and receivers, as well as statistical averaging over a large number of experimental trials. For the MISO configuration, accurate determination of $\langle E_b \rangle$ requires measuring $\langle V_{0,2} \rangle$, $\langle V_{1,2} \rangle$, and $\langle H_2 \rangle$, while the SIMO configuration requires the joint receivers to measure $\langle V_{0,2} \rangle$, $\langle V_{1,2} \rangle$, $\langle H_1 \rangle$, and $\langle H_2 \rangle$, since every term of the receiver region enters the conditional ground-state energy $\lambda_{\min}(H_\mu)$ and therefore the SIMO efficiency; in particular the inter-receiver bond $\langle V_{1,2} \rangle$ is retained as the positive cost of the joint operation rather than discarded. These measurements can only yield correct results when Alice, Charlie, and Bob communicate appropriately through the quantum circuits described in our protocol.

The distributions of quantum states obtained from \textit{ibm\_kyiv} are presented in Fig.~\ref{fig:miso_results} and Fig.~\ref{fig:simo_results}, where we compare the raw experimental results with those from \textit{qasm\_simulator}. Our analysis includes a comprehensive comparison between the 2-qubit minimal system and our extended 3-qubit system, examining both raw and error-mitigated measurement outcomes.

\subsection{Measurement Error Mitigation}

To address the impact of measurement errors inherent in current quantum hardware, we employed a measurement error mitigation technique at readout level-2. This method involves the following steps:

\begin{enumerate}
    \item We prepared a complete set of calibration circuits spanning the full computational basis of the Hilbert space (4 circuits for 2-qubit measurements, 8 circuits for 3-qubit measurements).
    \item These calibration circuits were executed immediately before the QET experiments to obtain the measurement error probability distributions characteristic of the device at that time.
    \item We constructed a calibration matrix from these probability distributions, which characterizes the measurement confusion between different computational basis states.
    \item This calibration matrix was then applied to correct the raw measurement outcomes from our QET circuits.
\end{enumerate}

The average measurement fidelity for each quantum computer is summarized in Table~\ref{tab:backend_comparison}. The histograms of observed quantum states exhibited consistent patterns across all quantum computers utilized in this study. Notably, the histograms generated from measurements of $H_2$ demonstrate excellent agreement with simulator results, confirming the high accuracy of our implementation. The enhancement of measurement accuracy through error mitigation is quantitatively verified by the results presented in Table~\ref{tab:simulation_results}.

The observation of negative energy expectation values is of paramount significance in this investigation, as it represents the definitive signature of successful quantum energy teleportation. While raw experimental data from quantum hardware initially showed discrepancies compared to ideal simulator results, the application of measurement error mitigation techniques substantially improved the agreement. Crucially, error mitigation enabled the observation of negative energy expectation values that align with theoretical predictions, thereby confirming the successful realization of the extended QET protocol on real quantum hardware. These results demonstrate that contemporary superconducting quantum computers possess sufficient coherence and gate fidelity to implement quantum energy teleportation protocols, paving the way for future applications in quantum communication and quantum thermodynamics.

\begin{table*}[!t]
    \centering
    \setlength{\extrarowheight}{1pt}
    \setlength{\tabcolsep}{4.5pt}
        \begin{tabular}{|c|c|c|c|c|c|c|}
    \hline
        Backend &  &  \multicolumn{1}{|c|}{2 Qubit QET} & \multicolumn{2}{|c|}{3 Qubit QET - MISO} & \multicolumn{2}{|c|}{3 Qubit QET - SIMO} \\
        \hline
        \hline
         &  & \((h, k) = (1.5, 1)\) & \((h, k) = (1, 4) \) & \((h, k) = (1, 3)\) &  \((h, k) = (1, 4) \) & \((h, k) = (1, 3)\)\\
        \hline
        \multicolumn{7}{|c|}{ \((E_a)\)} \\
        \hline
        Analytical value & & 1.2481 & 2.0551 & 1.7370 & 0.7700 & 0.7960 \\
        \hline
        \texttt{qasm\_simulator} &  & $1.2510 \pm 0.0048$ & $2.0420 \pm 0.0036$ & $1.7468 \pm 0.0048$ & $0.7684 \pm 0.0042$ & $0.7961 \pm 0.0041$ \\
        \hline
        \texttt{ibm\_brisbane} & error mitigated & $1.1486 \pm 0.0086$ & $1.8558 \pm 0.0065$ & $1.5844 \pm 0.0067$ & $0.7073 \pm 0.0053$ & $0.7561 \pm 0.0044$ \\
         & unmitigated & $0.8637 \pm 0.0110$ & $1.8493 \pm 0.0089$ & $1.5644 \pm 0.0095$ & $0.6542 \pm 0.0116$ & $0.7154 \pm 0.0091$ \\
        \hline
        \texttt{ibm\_sherbrooke} & error mitigated & $1.1846 \pm 0.0056$ & $1.9488 \pm 0.0052$ & $1.5741 \pm 0.0058$ & $0.7434 \pm 0.0084$ & $0.7420 \pm 0.0043$ \\
         & unmitigated & $0.9157 \pm 0.0088$ & $1.7494 \pm 0.0078$ & $1.2506 \pm 0.0078$ & $0.6950 \pm 0.0093$ & $0.6158 \pm 0.0059$ \\
        \hline
        \texttt{ibm\_kyiv} & error mitigated & $1.1843 \pm 0.0082$ & $1.8556 \pm 0.0050$ & $1.7420 \pm 0.0072$ & $0.7133 \pm 0.0088$ & $0.7916 \pm 0.0063$ \\
         & unmitigated & $1.0222 \pm 0.0066$ & $1.5305 \pm 0.0088$ & $1.4425 \pm 0.0061$ & $0.6292 \pm 0.0116$ & $0.6638 \pm 0.0076$ \\
        \hline
        \multicolumn{7}{|c|}{ \((E_c)\)} \\
        \hline
        Analytical value & &  & 2.0551 & 1.7370 &  &  \\
        \hline
        \texttt{qasm\_simulator} &  &  & $2.0583 \pm 0.0041$ & $1.7345 \pm 0.0041$ &  &  \\
        \hline
        \texttt{ibm\_brisbane} & error mitigated &  & $1.9550 \pm 0.0087$ & $1.7230 \pm 0.0059$ &  &  \\
         & unmitigated &  & $1.4616 \pm 0.0050$ & $1.2790 \pm 0.0083$ &  &  \\
        \hline
        \texttt{ibm\_sherbrooke} & error mitigated &  & $1.9980 \pm 0.0049$ & $1.7332 \pm 0.0068$ &  &  \\
         & unmitigated &  & $1.6138 \pm 0.0051$ & $1.5452 \pm 0.0094$ &  &  \\
        \hline
        \texttt{ibm\_kyiv} & error mitigated &  & $1.9752 \pm 0.0068$ & $1.6371 \pm 0.0063$ &  &  \\
         & unmitigated &  & $1.7441 \pm 0.0065$ & $1.5531 \pm 0.0082$ &  &  \\
        \hline
        \multicolumn{7}{|c|}{ \((V_{ab})\)} \\
        \hline
        Analytical value & & -0.4906 & -0.6992 & -0.5381 & -0.6140 & -0.4550 \\
        \hline
        \texttt{qasm\_simulator} &  & $-0.4921 \pm 0.0038$ & $-0.6979 \pm 0.0039$ & $-0.5410 \pm 0.0046$ & $-0.6104 \pm 0.0036$ & $-0.4552 \pm 0.0046$ \\
        \hline
        \texttt{ibm\_brisbane} & error mitigated & $-0.4091 \pm 0.0063$ & $-0.6876 \pm 0.0078$ & $-0.5230 \pm 0.0045$ & $-0.5832 \pm 0.0076$ & $-0.4556 \pm 0.0085$ \\
         & unmitigated & $-0.2737 \pm 0.0046$ & $-0.5665 \pm 0.0075$ & $-0.4205 \pm 0.0111$ & $-0.4152 \pm 0.0120$ & $-0.3960 \pm 0.0108$ \\
        \hline
        \texttt{ibm\_sherbrooke} & error mitigated & $-0.4469 \pm 0.0112$ & $-0.6654 \pm 0.0045$ & $-0.5156 \pm 0.0088$ & $-0.5801 \pm 0.0075$ & $-0.4548 \pm 0.0071$ \\
         & unmitigated & $-0.2229 \pm 0.0083$ & $-0.5179 \pm 0.0113$ & $-0.4698 \pm 0.0089$ & $-0.5200 \pm 0.0093$ & $-0.3889 \pm 0.0057$ \\
        \hline
        \texttt{ibm\_kyiv} & error mitigated & $-0.3804 \pm 0.0063$ & $-0.6417 \pm 0.0066$ & $-0.5298 \pm 0.0044$ & $-0.5498 \pm 0.0090$ & $-0.4400 \pm 0.0083$ \\
         & unmitigated & $-0.3089 \pm 0.0045$ & $-0.5773 \pm 0.0073$ & $-0.3683 \pm 0.0069$ & $-0.5548 \pm 0.0075$ & $-0.4037 \pm 0.0104$ \\
        \hline
        \multicolumn{7}{|c|}{ \((V_{bc})\)} \\
        \hline
        Analytical value & &  & -0.6994 & -0.5383 & 1.1380 & 0.8030 \\
        \hline
        \texttt{qasm\_simulator} &  &  & $-0.7032 \pm 0.0038$ & $-0.5371 \pm 0.0043$ & $1.1337 \pm 0.0037$ & $0.8108 \pm 0.0049$ \\
        \hline
        \texttt{ibm\_brisbane} & error mitigated &  & $-0.6293 \pm 0.0053$ & $-0.4886 \pm 0.0071$ & $1.1042 \pm 0.0083$ & $0.7858 \pm 0.0074$ \\
         & unmitigated &  & $-0.5613 \pm 0.0078$ & $-0.3935 \pm 0.0107$ & $0.9318 \pm 0.0055$ & $0.5796 \pm 0.0107$ \\
        \hline
        \texttt{ibm\_sherbrooke} & error mitigated &  & $-0.6558 \pm 0.0075$ & $-0.4918 \pm 0.0049$ & $1.0837 \pm 0.0052$ & $0.7840 \pm 0.0061$ \\
         & unmitigated &  & $-0.4934 \pm 0.0060$ & $-0.3710 \pm 0.0070$ & $0.8448 \pm 0.0106$ & $0.6206 \pm 0.0113$ \\
        \hline
        \texttt{ibm\_kyiv} & error mitigated &  & $-0.6689 \pm 0.0061$ & $-0.4979 \pm 0.0073$ & $1.1262 \pm 0.0051$ & $0.7667 \pm 0.0069$ \\
         & unmitigated &  & $-0.5449 \pm 0.0066$ & $-0.4029 \pm 0.0116$ & $0.9883 \pm 0.0104$ & $0.6844 \pm 0.0112$ \\
        \hline
        \multicolumn{7}{|c|}{ \((H_b)\)} \\
        \hline
        Analytical value & & 0.3480 & 0.2078 & 0.2135 & 0.0060 & 0.0160 \\
        \hline
        \texttt{qasm\_simulator} &  & $0.3487 \pm 0.0038$ & $0.2088 \pm 0.0037$ & $0.2105 \pm 0.0036$ & $0.0064 \pm 0.0040$ & $0.0151 \pm 0.0040$ \\
        \hline
        \texttt{ibm\_brisbane} & error mitigated & $0.3590 \pm 0.0047$ & $0.1932 \pm 0.0049$ & $0.1922 \pm 0.0078$ & $0.0075 \pm 0.0068$ & $0.0167 \pm 0.0054$ \\
         & unmitigated & $0.4302 \pm 0.0039$ & $0.1857 \pm 0.0119$ & $0.1793 \pm 0.0125$ & $0.0071 \pm 0.0080$ & $0.0121 \pm 0.0107$ \\
        \hline
        \texttt{ibm\_sherbrooke} & error mitigated & $0.3390 \pm 0.0084$ & $0.1813 \pm 0.0091$ & $0.2145 \pm 0.0054$ & $0.0171 \pm 0.0051$ & $0.0164 \pm 0.0077$ \\
         & unmitigated & $0.4871 \pm 0.0073$ & $0.1531 \pm 0.0093$ & $0.1595 \pm 0.0059$ & $-0.0126 \pm 0.0123$ & $0.0248 \pm 0.0105$ \\
        \hline
        \texttt{ibm\_kyiv} & error mitigated & $0.3559 \pm 0.0047$ & $0.1963 \pm 0.0052$ & $0.1931 \pm 0.0048$ & $-0.0025 \pm 0.0090$ & $0.0180 \pm 0.0080$ \\
         & unmitigated & $0.3737 \pm 0.0037$ & $0.2070 \pm 0.0108$ & $0.1448 \pm 0.0086$ & $0.0029 \pm 0.0087$ & $0.0081 \pm 0.0095$ \\
        \hline
    \end{tabular}\\
    \caption{Simulation results for different backends and methods compared to analytical values. For SIMO, $E_a$ is Alice's single-sender deposit and $E_c$ is blank (there is no second sender). For MISO, the combined Alice+Charlie deposit $\langle E_{AC}\rangle$ from the joint (entangle-then-measure) measurement is attributed symmetrically to the two senders, $E_a=E_c=\langle E_{AC}\rangle/2$. $V_{ab}=V_{0,2}$ and $V_{bc}=V_{1,2}$ are the two interaction bonds, the sender-receiver bond and the inter-receiver bond respectively. For MISO both are extracted ($V_{ab}=V_{bc}<0$). For SIMO the joint receivers act on $V_{ab}$ and $V_{bc}$ together: at the joint optimum the sender-receiver bond is extracted ($V_{ab}<0$) while the inter-receiver bond rises ($V_{bc}>0$), and this positive cost is fully included in the work balance. $H_b=\langle H_2\rangle$ is Bob's local-field energy. The SIMO efficiency is $\eta_{\mathrm{SIMO}}=W/E_0$ with $W=E_0-\sum_\mu p_\mu\lambda_{\min}(H_\mu)$, so all three returned quantities ($V_{ab}$, $V_{bc}$, $H_b$) enter through the conditional ground-state energy of the joint receiver pair. Backend entries are reported as mean $\pm$ standard error over the measurement shots; unmitigated readings are biased toward zero by decoherence, while error mitigation restores them close to the analytical values.}
    \label{tab:simulation_results}
\end{table*}

\begin{table*}[!t]
\centering
\begin{tabular}{|c|c|c|c|c|c|}
\hline
\textbf{Backend} & \textbf{Qubit} & \textbf{T1 (us)} & \textbf{T2 (us)} & \textbf{Frequency (GHz)} & \textbf{Readout Error} \\ \hline
 \multirow{4}{*}{ibm\_kyiv}      & 0  & 240.6   & 300.7   & 4.656 & 6.800e-3  \\ 
                                & 1  & 456.53  & 210.99  & 4.535 & 2.800e-3  \\ 
                                & 2  & 118.61  & 87.21   & 4.68  & 5.900e-3  \\ 
                                & 3  & 253.51  & 159.27  & 4.607 & 5.800e-3  \\ \hline
 \multirow{4}{*}{ibm\_sherbrooke} & 0  & 427.97  & 107.97  & 4.636 & 1.580e-2  \\ 
                                & 1  & 310.73  & 325.86  & 4.736 & 1.820e-2  \\ 
                                & 2  & 265.62  & 187.5   & 4.819 & 1.820e-2  \\ 
                                & 3  & 315.87  & 171.31  & 4.747 & 1.180e-2  \\ \hline
 \multirow{4}{*}{ibm\_brisbane}   & 10 & 325.43  & 292.43  & 4.832 & 1.550e-2  \\ 
                                & 11 & 354.78  & 295.86  & 4.972 & 1.343e-1  \\ 
                                & 12 & 350     & 158.02  & 4.934 & 1.650e-2  \\ 
                                & 13 & 307.89  & 129.35  & 5.006 & 1.230e-2  \\ \hline
\end{tabular}

\caption{Calibration Data of IBM Quantum Systems (Kyiv, sherbrooke, brisbane)}
\label{tab:calibration_data}
\end{table*}

\begin{table*}[!t]
\centering

\begin{tabular}{|c|c|c|c|}
\hline
\textbf{}              & \textbf{ibm\_sherbrooke} & \textbf{ibm\_kyiv} & \textbf{ibm\_brisbane} \\ \hline
\textbf{Qubits}            & 127                      & 127                 & 127                     \\ 
\textbf{2Q Error (best)}    & 2.35e-3                  & 3.93e-3             & 3.28e-3                 \\ 
\textbf{2Q Error (layered)} & 1.63e-2                  & 1.50e-2             & 2.77e-2                 \\ 
\textbf{CLOPS}             & 30K                      & 30K                 & 30K                     \\ 
\textbf{Median SX error}    & 2.502e-4                 & 2.629e-4            & 2.331e-4                \\ 
\textbf{Median Readout error} & 1.300e-2                & 8.600e-3            & 1.410e-2                \\ 
\textbf{Median ECR error}   & 7.697e-3                 & 1.166e-2            & 7.877e-3                \\ 
\textbf{Median T1 (us)}     & 247.83                   & 263.59              & 225.19                  \\
\textbf{Median T2 (us)}     & 179.86                   & 121.37              & 151     \\   
\textbf{Iterations}     & 1024                   & 1024              & 1024
\\ \hline
\end{tabular}

\caption{Comparison of IBM Quantum Backends}
\label{tab:backend_comparison}
\end{table*}

It was established earlier in \cite{key14} that  we have observed for all parameter \( (k, h) \) combinations, negative \( \langle V \rangle \) and it is true for all type of quantum computers of IBM.Previously which was proved for 2-qubit system in \cite{key14}, is also true for our 3-qubit system. Bob can extract greater energy if only \( V_{0,2} \) and \( V_{1,2} \)is observed, since \( \langle H_2 \rangle \) is always positive (Fig.~\ref{fig:miso_results}). For practical purposes using minimal model was enough as said by the author of the correspond paper, in addition, our model performs even better, which in turn takes QET one step forward. Either way, we have to keep in mind that Bob's energy becomes smaller when he observes \( H_2 \).

\section{Some details of the model}

 \justifying{This section is comprised of a comprehensive description of the model utilized in our study. Additional information are available in Hotta's original papers. It is crucial to acknowledge that the lowest energy state of the whole Hamiltonian is not the lowest energy state of local operators.}

\begin{figure*}[!t]
    \centering
    \begin{subfigure}[b]{0.45\textwidth}
        \includegraphics[width=\textwidth]{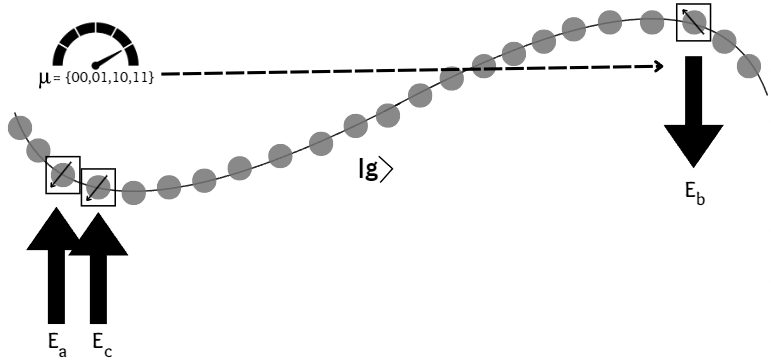}
 
    \end{subfigure}
    \begin{subfigure}[b]{0.45\textwidth}
        \includegraphics[width=\textwidth]{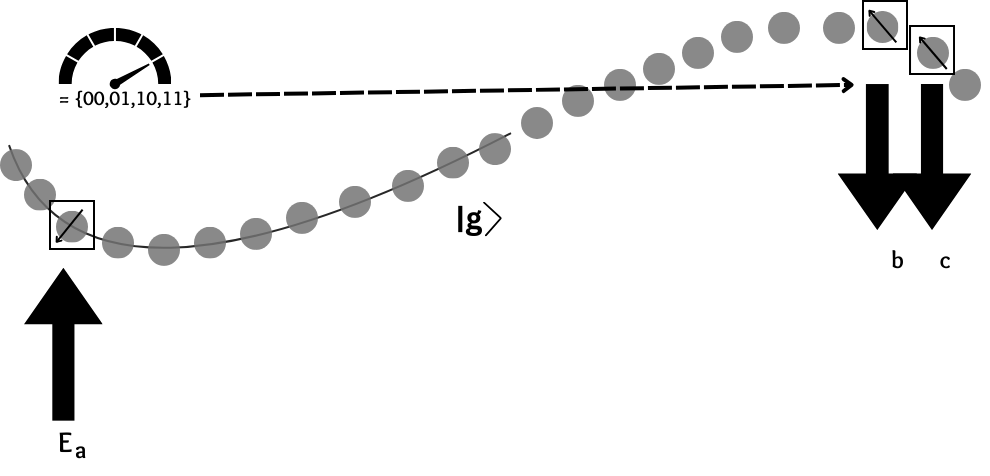}
        
    \end{subfigure}
    \caption{Spin Chain Diagram Of Enhanced QET Model ( MISO \& SIMO )}
    \label{fig:spin_chain}
\end{figure*}

 If we were to discuss enhanced QET for spin chain systems, we have to concentrate on short time scales, in which time evolution of the Hamiltonian of the spin chain is negligible. Further, we can also assume the nonrelativistic limit that LOCC for the spins 
can be repeated many times even in a short time interval. According to the diagram in Fig.~\ref{fig:spin_chain}, Alice, Charlie and Bob share many near-critical spin chains in the ground state $\lvert g \rangle$, which is entangled and has a large correlation length $l$. Alice is situated at site $n_A$, Charlie at site $n_C$ and Bob at site $n_B$. Alice and Charlie share near distance , however Bob is in a good distance from them:\\
In case of MISO, 
\begin{align*}
\lvert n_A - n_B \rvert \sim O(l) \gg 1.\\
\lvert n_C - n_B \rvert \sim O(l) \gg 1.
\end{align*}
In case of SIMO, 
\begin{align*}
\lvert n_A - n_B \rvert \sim O(l) \gg 1.\\
\lvert n_A - n_c \rvert \sim O(l) \gg 1.
\end{align*}

 In order to comprehend the non-triviality of the QET protocol, it is crucial to acknowledge that regardless of the specific unitary operations $W_1$ \& $W_2$ applied to Bob's qubit following Alice's and Charlie's measurements, it is impossible to extract any energy without classical communication \cite{key18,key22}. This can be verified in the case of the MISO model by,
\begin{equation}
\text{Tr}[\rho_W H_{tot}] - ( \langle E_0 \rangle + \langle E_1 \rangle ) = \langle g| W_2^\dagger W_1^\dagger H_{tot} W_1 W_2 |g\rangle \geq 0
\label{eq:pw1}
\end{equation}
where
\begin{equation}
\begin{aligned}
\rho_W = W_1 W_2 \Big( \sum_{\mu_0, \mu_1 = \pm 1} & P_1(\mu_1) P_0(\mu_0) |g\rangle \langle g| \\
& \times P_0(\mu_0) P_1(\mu_1) \Big) W_2^\dagger W_1^\dagger
\end{aligned}
\end{equation}

In the case of SIMO, the joint receivers apply their operation $W_1 W_2$ to Charlie's and Bob's qubits following only Alice's measurement.

\begin{equation}
\text{Tr}[\rho_W H_{tot}] - \langle E_0 \rangle = \langle g| W_2^\dagger W_1^\dagger H_{tot} W_1 W_2 |g\rangle \geq 0
\label{eq:pw2}
\end{equation}
where
\begin{equation}
\rho_W =  W_1 W_2 \left( \sum_{\mu = \pm 1}  P_0(\mu) |g\rangle \langle g| P_0(\mu)  \right) W_2^\dagger W_1^\dagger
\end{equation}
The inequalities \eqref{eq:pw1} and \eqref{eq:pw2} hold because the fixed (measurement-independent) unitaries $W_1, W_2$ commute with the projective measurements acting on the complementary qubits, and the ground state $|g\rangle$ has zero energy by construction [Eq.~\eqref{eq:zero_energy_all}], so the right-hand side is the mean energy of a state evaluated against a Hamiltonian whose spectrum is non-negative. Energy extraction therefore strictly requires the conditional dependence of the receivers' operations on the classically communicated outcomes.

Bob's local system's time evolution, if he does not perform any operations on his own system after Alice's measurement, is depicted as
\begin{equation}
\langle H_i(t) \rangle = \text{Tr}[\rho_M e^{i t H} H_i e^{-i t H}]
\end{equation}
\begin{equation}
\langle V(t) \rangle = \text{Tr}[\rho_M e^{i t H} V_{i,j} e^{-i t H}] = 0, \{i,j\} = [0,2] \& i<j
\end{equation}
where $\rho_M = \sum_{\mu = \pm 1} P_1(\mu_1) P_0(\mu_0) |g\rangle \langle g| P_0(\mu_0)  P_1(\mu_1)$\\

The time evolution of the system, results in energy transfer to Bob's local system. However, this transfer is simply the propagation of energy in the usual manner. In Quantum Energy Teleportation (QET), energy is not acquired through the natural progression of time inside the system, but rather immediately through communication. Given that we are examining a non-relativistic quantum many-body system, the rate at which energy travels is significantly slower than the speed of light. Optical communication, a kind of classical communication, may transmit information to distant locations at a significantly faster rate than the temporal progression of physical systems. As, QET is recognized as a rapid energy transmission protocol, our method employ efficiency on top of that.

\section{Thermodynamic Bounds and Entropy Calculation}
\label{sec:entropy_bound}

To rigorously evaluate the efficiency of the 3-qubit Quantum Energy Teleportation (QET) protocol, we analyze the information-energy relation. As established in the minimal QET model \cite{key14}, the extraction of energy by the receiver is strictly bounded by the consumption of ground-state entanglement, quantified by the decrease in mutual information between the subsystems.

\subsection{Thermodynamic Justification}

Both protocols are consistent with thermodynamics, and their mechanism is the same as in the minimal QET model \cite{key14,key18}. The system begins in the ground state $|g\rangle$, whose energy we have shifted to zero, $\langle g|H_{\text{tot}}|g\rangle = 0$. A projective measurement on the sender side is \emph{not} a unitary operation: it dephases the ground state in the measured basis, and because the measured observable does not commute with the local field, it raises the mean energy to a strictly positive value. This deposited energy is supplied by the measurement apparatus, since no energy is created from nothing; the apparatus does work on the system.

The receiver then applies a \emph{local} unitary conditioned on the classical outcomes (a LOCC operation); no energy is physically transported. Because the map from $|g\rangle$ to the post-measurement ensemble is non-unitary, the receiver's local energy $\langle H_2 + V_{0,2} + V_{1,2}\rangle$ can be driven \emph{below} its ground-state value of zero. This negative local energy density is the thermodynamic signature of QET. Global energy conservation is never violated: the negative energy retrieved by the receiver is balanced by the positive energy deposited at the sender(s), so the extracted work obeys $W \le E_{\text{dep}}$. For the MISO protocol, part of the deposit is placed by the entangler directly into the receiver's bond; subtracting that part, the net negative energy the receiver extracts relative to the senders' deposit gives $\eta_{\text{MISO}} = -\langle E_b\rangle/\langle E_S\rangle \approx 34\%$ to $42\%$, with conservation respected throughout.

In the SIMO protocol, a single sender (Alice) deposits $E_0$ and the two receivers Bob and Charlie jointly extract negative energy. Alice's projective measurement is non-unitary and supplies $E_0$ from the apparatus; because $X_0$ commutes with every receiver term [Eq.~\eqref{eq:commute}], the deposit injects no energy directly into the receiver pair, which instead lower their joint energy below its ground-state value by a conditional two-qubit rotation. The inter-receiver bond $V_{1,2}$ is fully retained in this balance, so the extracted work obeys $W \le E_0$ and is honest, and the negative local energy density at the receivers is the thermodynamic signature of the protocol \cite{ABN}. In all cases the extractable energy remains bounded above by the consumption of ground-state correlations, the energy-entropy bound derived in the following subsections, so no protocol extracts more than the entanglement structure of $|g\rangle$ permits.

\subsection{Ground State and Subsystem Entropy}
The total system is initialized in the ground state $|g\rangle$ of the Hamiltonian $H_{tot}$. Unlike the 2-qubit singlet state used in the minimal model, our 3-qubit ground state is defined by the single coefficient $M$, which satisfies the zero-energy constraints derived in Section II. The properly normalized ground state is:
\begin{equation}
    |g\rangle = \frac{1}{\mathcal{N}}\bigl(|111\rangle - M\,(|100\rangle + |010\rangle + |001\rangle)\bigr),
    \label{eq:gs_normalized}
\end{equation}
where the normalization constant is
\begin{equation}
    \mathcal{N} = \sqrt{1 + 3M^2}.
    \label{eq:gs_norm_const}
\end{equation}
The coefficient $M$ is a real-valued parameter determined by $h$ and $k$ through Eq.~\eqref{eq:M}, and this state is identical to the ground state introduced in Eq.~\eqref{eq:ground_state}. For any physically admissible choice $h,k>0$, $M$ is non-zero, so $\mathcal{N}>1$. The limiting case $\mathcal{N}=1$ (i.e., $M=0$) would correspond to the trivial product state $|111\rangle$, which carries no entanglement and for which QET is inoperative; the non-trivial QET regime is therefore precisely the regime $\mathcal{N}>1$. Every quantity of the form $\langle g|O|g\rangle$ appearing in the manuscript is evaluated using the normalized state \eqref{eq:gs_normalized}, i.e.,
\begin{multline}
    \langle g|O|g\rangle \equiv \frac{1}{\mathcal{N}^2}\bigl(\langle 111| - M\,(\langle 100| + \langle 010| + \langle 001|)\bigr) \\
    \times\, O\,\bigl(|111\rangle - M\,(|100\rangle + |010\rangle + |001\rangle)\bigr).
    \label{eq:gs_expval}
\end{multline}
The zero-mean-energy conditions~\eqref{eq:zero_energy_all} that fix the constants~\eqref{eq:constants} are derived and verified using this normalized form. The correlations in this state provide the resource for energy teleportation. We consider the bipartition between the sender (Alice, subsystem $A$) and the receivers (Bob and Charlie, subsystem $BC$). The reduced density matrix of the receiver's subsystem prior to measurement is obtained by tracing out Alice's qubit:
\begin{equation}
    \rho_{BC} = \text{Tr}_A \left( |g\rangle\langle g| \right).
\end{equation}
The initial Von Neumann entropy of the receiver's subsystem is $S_{BC} = -\text{Tr}(\rho_{BC} \ln \rho_{BC})$.

\subsection{Measurement and Entropy Shift}
In the protocol, Alice performs a projective measurement on qubit $A$ using the operator $P_0(\mu) = \frac{1}{2}(1 + \mu X_0)$, obtaining outcome $\mu \in \{+1, -1\}$. This measurement collapses the global state and alters the entropy of the remaining system. The post-measurement average entropy is given by:
\begin{equation}
    S_{avg} = \sum_{\mu \in \{+1, -1\}} p_\mu S(\rho_{BC}^{(\mu)}),
\end{equation}
where $p_\mu = \langle g | P_0(\mu) | g \rangle$ is the probability of outcome $\mu$, and $\rho_{BC}^{(\mu)}$ is the reduced density matrix of the receivers conditioned on Alice's result.

The net information cost of the protocol is the difference between the initial entanglement entropy and the average post-measurement entropy:
\begin{equation}
    \Delta S_{A:BC} = S_{BC} - \sum_{\mu=\pm 1} p_\mu S_{BC}(\mu).
    \label{eq:entropy_diff}
\end{equation}

\subsection{Energy-Entropy Bound}
Following the thermodynamic principles of QET \cite{key25}, the maximum energy extractable by the receiver, $E_{extracted}$, is bounded by this entropy shift. Generalizing the bound derived in the minimal model (Eq. A11 in \cite{key14}) to our 3-qubit protocols, the relevant energy scale is no longer $\sqrt{h^2+k^2}$ but the eigenvalue $K=\sqrt{h^2+hk+k^2}$ that fixes the 3-qubit ground state (Section II), so that
\begin{equation}
    \Delta S_{A:BC} \ge \beta \frac{E_{extracted}}{\sqrt{h^2 + hk + k^2}},
\end{equation}
where $\beta$ is a coefficient determined by the ground-state parameter $M$ and the coupling strength. The bound applies to both the MISO and the SIMO protocols: in each, the work Bob extracts cannot exceed what the consumed mutual information $\Delta S_{A:BC}$ permits. The higher extraction efficiency of the 3-qubit protocols corresponds to the enhanced $\Delta S_{A:BC}$ available in the 3-qubit Ising ground state, which raises the energy a receiver can extract, while the exactly computable single-receiver ergotropy bound of Section~\ref{subsec:joint_bound} supplies the matching upper limit on the work itself. This confirms that the additional qubit allows for a greater consumption of mutual information, thereby raising the theoretical ceiling for energy teleportation.

\subsection{Optimality of the Receiver's Conditional Rotation}
\label{subsec:joint_bound}

A complementary, exactly computable bound confirms that the receiver's simple conditional rotation is \emph{optimal}. After the combined (entangle-then-measure) measurement of the senders, qubits $0$ and $1$ are projected onto definite $X$ eigenstates, fully disentangled from Bob: $P_0(\mu_0)P_1(\mu_1)\,C\,|g\rangle \propto |\mu_0,\mu_1\rangle \otimes |\chi_{\mu_0\mu_1}\rangle$. The only remaining freedom is a unitary on Bob's single qubit, which is governed by the \emph{conditional single-qubit Hamiltonian}
\begin{equation}
    H_{\mu_0\mu_1} = \langle \mu_0,\mu_1 |\, C^\dagger H_{\text{tot}} C\, |\mu_0,\mu_1\rangle ,
    \label{eq:cond_H}
\end{equation}
a positive-semidefinite $2\times2$ operator obtained by contracting the projected sender qubits. No operation on Bob's qubit can lower the system energy below $\lambda_{\min}(H_{\mu_0\mu_1})$, so the extractable work is bounded by
\begin{equation}
    W \;\le\; W_{\max} \;=\; \langle E_{AC}\rangle - \sum_{\mu_0,\mu_1} p_{\mu_0\mu_1}\, \lambda_{\min}(H_{\mu_0\mu_1}),
    \label{eq:joint_bound}
\end{equation}
the single-qubit ergotropy of Bob's conditional state \cite{ABN}. Because $H_{\mu_0\mu_1}$ and $|\chi_{\mu_0\mu_1}\rangle$ are real, the energy-minimising state lies in the $X$-$Z$ plane and is reached by a rotation about $Y$; Bob's conditional rotation $R_Y(2\phi)$ therefore \emph{saturates} the bound, and no more elaborate gate extracts more energy. As a consistency check, the corresponding bound for the minimal 2-qubit model is saturated exactly by Hotta's conditional rotation, $E_A - \lambda_{\min}(\langle\mu|H_{\text{tot}}|\mu\rangle) = \frac{1}{2}\bigl[\sqrt{\xi^2+\eta^2}-\xi\bigr]$, which our exact computation confirms ($W = 0.1425$ at $(h,k)=(1.5,1)$).

Table~\ref{tab:joint} reports the resulting MISO net efficiency $\eta_{\text{MISO}}=-\langle E_b\rangle/\langle E_S\rangle$ across the coupling range. Because the entangler-deposited part $\langle E_b^{(0)}\rangle$ is fixed by the deposit, maximising the work $W$ also maximises the net extraction $-\langle E_b\rangle$; the value obtained with Bob's single $R_Y(2\phi)$ rotation therefore equals the full single-qubit bound \eqref{eq:joint_bound} to numerical precision, confirming optimality. The efficiency depends only on the ratio $k/h$ and rises monotonically with it, from $23.7\%$ at $k/h=2$ to $42.2\%$ at $k/h=4$.

\begin{table}[!t]
\centering
\begin{tabular}{|c|c|c|}
\hline
$k/h$ & $\eta_{\text{MISO}}=-\langle E_b\rangle/\langle E_S\rangle$ & optimal (bound) \\ \hline
2   & 23.7\% & 23.7\% \\
2.5 & 29.2\% & 29.2\% \\
3   & 34.1\% & 34.1\% \\
3.5 & 38.4\% & 38.4\% \\
4   & 42.2\% & 42.2\% \\ \hline
\end{tabular}
\caption{MISO net teleportation efficiency $\eta_{\text{MISO}}=-\langle E_b\rangle/\langle E_S\rangle$ (the negative energy Bob extracts, after removing the entangler-deposited part, relative to the sender deposit) obtained with Bob's single conditional rotation $R_Y(2\phi)$, compared with the optimal single-qubit value. The two coincide, confirming that the simple local rotation is optimal. Values are exact and depend only on $k/h$.}
\label{tab:joint}
\end{table}

This optimality also clarifies the roles of the parties: in MISO the two senders' combined measurement is what deposits the extractable energy into Bob's region, and a single local rotation by Bob then recovers all of it that the conditional state permits. The SIMO protocol obeys the analogous joint bound: after Alice's measurement the two receivers' optimal joint rotation reaches the ground state of the conditional Hamiltonian $H_\mu$ [Eq.~\eqref{eq:cond_H_simo}], so their joint extraction is likewise optimal.

\section{Implications for Real-World Applications}

Our experimental demonstration of the extended 3-qubit QET protocol with significantly enhanced energy extraction efficiency has profound implications for the development of quantum communication technologies across multiple timeframes, from near-term implementations to long-term quantum network infrastructures.

\subsection{Near-Term Applications and Experimental Feasibility}

It is crucial to recognize that, analogous to quantum state teleportation, quantum energy teleportation operates exclusively through Local Operations and Classical Communication (LOCC), without requiring the physical transport of energy carriers. This fundamental characteristic makes QET immediately implementable on current quantum computing platforms. Our extended QET model, requiring only 3 qubits with a circuit depth of approximately 9 gates, is well within the capabilities of contemporary quantum processors. This modest resource requirement enables immediate experimental validation in small-scale laboratory facilities using existing quantum computing and communication infrastructure.

The performance demonstrated by our protocol represents a substantial advancement toward practical quantum energy applications: in the MISO configuration two senders and a single receiver achieve a net teleportation efficiency of $34\%$ to $42\%$, well above the minimal 2-qubit system \cite{key14}, while in the SIMO configuration a single sender distributes energy to two joint receivers at an honest efficiency of $8\%$ to $10\%$, comparable to the 2-qubit value but realized over a multi-receiver geometry with every interaction term, including the inter-receiver bond, retained in the accounting. This improvement suggests promising directions for near-term applications in quantum memory systems \cite{key24,key25,key26,key27}, where efficient energy management is critical for maintaining coherence and fidelity. Moreover, our work addresses the crucial challenge of validating QET principles across diverse quantum systems and materials beyond the minimal model, thereby establishing a foundation for scalable implementations.

\subsection{Advanced Quantum Communication Protocols}

The enhanced capabilities of our 3-qubit QET protocol create opportunities for implementing sophisticated quantum communication schemes with improved efficiency and security. Emerging concepts such as Quantum Oblivious Transfer (QOT) \cite{Santos_OT}, quantum blockchain architectures, and Quantum Interactive Proofs \cite{Ikeda_ip} can potentially benefit from the multi-party energy teleportation framework established by our MISO and SIMO configurations. The ability to distribute a single sender's deposit to two joint receivers (SIMO) or to aggregate energy from multiple senders to a single receiver (MISO) provides architectural flexibility that may enhance the efficiency and security of these advanced protocols.

Furthermore, the successful demonstration of negative energy extraction through projective measurements and conditional operations validates fundamental principles underlying quantum resource theories. This validation extends beyond energy teleportation to inform the design of quantum batteries \cite{Ikeda_ip}, quantum thermal machines, and other quantum thermodynamic devices where controlled energy manipulation at the quantum level is essential.

\subsection{Long-Range Quantum Networks and Global Infrastructure}

The theoretical framework for unlimited-distance quantum energy teleportation has been established \cite{key27}, and our enhanced protocol brings this vision closer to practical realization. The capacity to transmit quantum energy across arbitrary distances at the speed of light limited only by classical communication will catalyze a transformative revolution in quantum communication technology. This capability enables a future wherein physical quantities, not merely quantum information, can be transmitted instantaneously to distant locations across a global-scale Quantum Internet accessible to end users.

Significant progress toward realizing quantum network infrastructure is already underway, with several pioneering quantum networks operational worldwide \cite{key29,key30,key31}, including the notable long-distance ($\sim$158 km) quantum network connecting sites in Long Island, New York \cite{key28}. The integration of QET protocols into these emerging quantum networks is anticipated to be achievable by the late 2020s, representing a critical milestone in the progression toward establishing QET on a planetary scale. Our extended protocol, with its demonstrated efficiency improvements and multi-party capabilities, advances this timeline by providing a more practical and scalable implementation framework.

\subsection{Quantum Information Economics and Market Implications}

The implementation of extended long-range QET will generate significant ramifications that transcend purely technological advances in information and communication systems. Both information and energy possess not only physical properties but also intrinsic economic value. The ability to trade physical quantities specifically, extractable energy directly over quantum networks will catalyze the emergence of entirely new economic markets and business models \cite{key32}.

In a quantum marketplace characterized by multiple entangled networks connecting numerous participants, receivers can dynamically select among competing senders based on optimization criteria including energy extraction efficiency, transmission fidelity, communication latency, and transaction costs \cite{key33,key34,key35,key36,key37}. This creates a competitive environment wherein the quality of entanglement, the efficiency of LOCC protocols, and the reliability of quantum hardware become quantifiable economic assets.

The concept of quantum information economics currently in its infancy will gain substantial significance as QET and related quantum communication technologies mature toward commercial deployment. The Hermitian operators' expectation values, conventionally interpreted as physical observables such as energy, acquire dual roles as both physical quantities and tradeable commodities. Teleported energy can serve conventional purposes (e.g., charging quantum batteries, powering quantum devices) or function as a medium of exchange, a store of value, or a unit of account within quantum economic systems.

Moreover, the multi-party configurations demonstrated in our work (MISO and SIMO) naturally extend to complex network topologies involving numerous participants. Game-theoretic considerations arise when multiple parties compete or cooperate to maximize energy extraction, minimize resource expenditure, or optimize other objective functions. The strategic interactions among participants in quantum energy markets introduce fascinating questions at the intersection of quantum physics, information theory, economics, and computational complexity theory.

\subsection{Fundamental Physics and Future Directions}

Beyond practical applications, our extended QET protocol provides new experimental tools for investigating fundamental questions in quantum physics. The ability to extract negative ground state energy from local and semi-local Hamiltonians through entanglement-assisted protocols offers insights into the nature of quantum correlations, the role of measurement in quantum mechanics, and the interplay between quantum information and thermodynamics.

The SIMO configuration, in particular, demonstrates that the energy from a single local deposit can be jointly extracted as negative energy density by two distant receivers, challenging classical intuitions about energy locality. Understanding the limits and capabilities of such multi-party quantum protocols may illuminate fundamental aspects of quantum many-body physics, including quantum phase transitions, topological order, and emergent phenomena in strongly correlated systems.

Future research directions include: (1) extending the protocol to larger qubit numbers with more complex network topologies, (2) implementing QET in diverse physical platforms beyond superconducting qubits (e.g., trapped ions, photonic systems, neutral atoms), (3) integrating QET with quantum error correction to achieve fault-tolerant energy teleportation, (4) exploring connections between QET and holographic principles in quantum gravity, and (5) developing practical applications in quantum computing, quantum sensing, and quantum communication that leverage the unique capabilities of energy teleportation protocols.

\section{Conclusion} 

The ground state of a many-body quantum system is \mbox{usually} entangled which can be subjected to various \mbox{interesting} protocols and applications. Although the measurement of a subsystem destroys the entanglement, some energies are injected into the local system, some of which can be \mbox{retrieved} using local operations and classical communication. The minimal Quantum Energy Teleportation (QET) model exploits this phenomenon for two-qubit systems.  
This paper extends the minimal QET circuit to a 3-qubit system whose entangled ground state, prepared on a fully connected transverse-field Ising Hamiltonian [Eq.~\eqref{eq:Htot_def}], is shared among three protocol users. In the MISO configuration this overcomes the \mbox{limitation} of the low-energy extraction problem, the single receiver attaining a net teleportation efficiency of $34\%$ to $42\%$ from two combined senders; in the SIMO configuration a single sender distributes energy to two joint receivers at an honest efficiency of $8\%$ to $10\%$ (from the measured fields $H_1,H_2$ and bonds $V_{0,2},V_{1,2}$), comparable to the minimal 2-qubit model but realized over a multi-receiver geometry with no interaction term excluded from the accounting. This highlights the efficacy of our novel Ising Model Hamiltonian based on 3-qubit time-evolution energy dynamics and enables exploring non-trivial topological characteristics, novel symmetry groups, robust fault tolerance and better approximations of quantum fields. These achievements make the proposed protocol more efficient, with applications in quantum energy teleportation and condensed matter physics. 

\section{Acknowledgement}
 
We thank Mr. Syed Emad Uddin Shubha, Mr. Sabir Md Sanaullah and Mr. Md Zubair for fruitful communication and initial discussions. We acknowledge the use of
IBM quantum computers and Quantum Environment. M.R.C. Mahdy acknowledges the support of NSU internal grant and CTRGC grant 2024-25 and 2025-26.

\section{Competing interests}
 
The author declares that there is no competing financial interests.

\end{document}